\documentstyle[12pt]{article}
\newcommand{\beq}{\begin{equation}}
\newcommand{\eeq}{\end{equation}}
\begin{document}

\begin{titlepage}
\begin{center}
{\hbox to\hsize{hep-th/9810155  \hfill   MIT-CTP-2788}} 

{\hbox to\hsize{ 
\hfill PUPT-1815}}

{\hbox to\hsize{ 
\hfill BUHEP-98-26}}

\bigskip
\vspace{3\baselineskip}

{\Large \bf  
 
Out Of This World Supersymmetry Breaking\\}

\bigskip

\bigskip

{\bf  Lisa Randall}\\
\smallskip

{ \small \it  
Joseph Henry Laboratories,
Princeton University,
Princeton, NJ 08543, USA\\

and

 Center for Theoretical Physics,

Massachusetts Institute of Technology, Cambridge, MA 02139, USA }

\medskip
{\bf Raman Sundrum}\\
\smallskip
{\small \it Department of Physics, 

Boston University, Boston, MA 02215, USA}

\bigskip

{\tt  randall@baxter.mit.edu} \\
{\tt sundrum@budoe.bu.edu}

\bigskip

\vspace*{.5cm}

{\bf Abstract}\\
\end{center}
\noindent
We show that in a general  hidden sector model, supersymmetry
breaking necessarily generates at one-loop a  scalar and
gaugino mass as a consequence of the super-Weyl
anomaly. We study a scenario in which this
contribution dominates.
We consider the  Standard Model particles to be  localized
on a (3+1)-dimensional subspace or ``3-brane'' of a higher dimensional
spacetime, while supersymmetry breaking occurs off the 3-brane, either in the 
bulk
 or on another 3-brane. At least one extra
dimension is assumed   to be
 compactified roughly one to two orders of magnitude  below the
 four-dimensional Planck
 scale.  This framework is phenomenologically very
 attractive;  it introduces new possibilities for solving 
 the supersymmetric flavor problem, the gaugino mass problem, 
 the supersymmetric CP problem, and the  $\mu$-problem.
 Furthermore, the compactification  scale can be  
consistent with a unification of  gauge and gravitational couplings.
 We demonstrate these claims in  a four-dimensional effective theory below the
 compactification scale that  incorporates the  relevant features of the 
underlying higher dimensional theory and  the contribution of the super-Weyl 
anomaly.
Naturalness constraints follow not only from symmetries
but also from the higher dimensional origins of the theory.  We also introduce
 additional bulk contributions to the MSSM soft masses. This scenario is
 very predictive: the  gaugino masses,  squark masses, and $A$ terms
 are given in terms of MSSM renormalization group functions.

\bigskip

\bigskip

\end{titlepage}

\section{Introduction}
If string theory is correct,  fundamental physics is
higher dimensional and has a large amount of supersymmetry.
Because we live in four dimensions with no supersymmetry,
something nontrivial must necessarily happen with the other dimensions,
which are compactified or truncated with boundaries.  In general,
reducing the dimensionality can reduce the amount of supersymmetry.
In fact, for extended supersymmetry, the breaking is necessarily
of a geometric nature \cite{ss}.  For $N=1$ supersymmetry however,
the breaking can be effectively four-dimensional and field theoretical, 
independent of the geometry of the small compact dimensions.

In this paper, 
we investigate  the not so radical assumption (in light
of the above) that the mechanism for
 breaking  the final $N=1$ supersymmetry of the
MSSM  is also
intimately tied to the geometry of the extra dimensions. We assume 
(for reasons we will later explain) that 
the MSSM itself is confined to a (3+1)-dimensional subspace, or ``3-brane'',
of the higher dimensional spacetime. Supersymmetry
is then broken somewhere else.
Known possible mechanisms  include Scherk-Schwarz
compactification\cite{ss} \cite{ss1} \cite{ss2},
rotated branes \cite{rot1} \cite{rot2} \cite{rot3}, 
and gaugino condensation or dynamical supersymmetry
breaking  on a separate 3-brane along the lines
of \cite{hw,horava}. They all share an important common
feature: supersymmetry is preserved
locally on the visible sector 3-brane, but
is broken globally. In the decompactification limit,
supersymmetry breaking would disappear from the visible world.
We refer to such supersymmetry breaking  as having originated in
 a ``sequestered sector''.

What this means from the vantage point of four dimensions
is that the ultraviolet convergence is improved.  Indeed the
supersymmetry-breaking effects we compute are finite, cut off by 
the compactification
scale. This is remarkable given the well known non-renormalizability and
ultraviolet sensitivity of
(super)gravity, especially in higher dimensions.  But because
supersymmetry breaking in the visible world requires
communication across the extra dimensions, the diagrams
contributing to supersymmetry-violating effects are
 {\it finite}. This was previously 
pointed out in
the context of Scherk-Schwarz supersymmetry breaking in
Refs. \cite{gc2} \cite{dimopoulos}, and in the toy model of
Ref. \cite{peskin}. This gives a starting point, in 
the context of field theory, for
solving the supersymmetric flavor problem.  Generally, the 
problem involves the fact that there are arbitrary flavor dependent
higher dimension operators in the Kahler potential which upon
supersymmetry breaking give flavor dependent masses. In fact, upon
running the squark masses using known flavor violation, one finds
a logarithmic dependence on the renormalization scale, indicating
the necessity for such counterterms. It could be that there is a definite
scale at which the masses are degenerate related to the string scale,
and string ``miracles'' solve the flavor problem. However,
it is much more compelling to see the solution in the
low-energy field theory. We argue that sequestered supersymmetry
breaking eliminates the necessity for such counterterms. Essentially
the higher dimensional theory generalizes the notion of naturalness;
the absence of counterterms might not be obvious from the standpoint
of symmetries of the low-energy theory, but can nonetheless
be guaranteed by assumptions in the higher dimensional theory.

We will show                          that supersymmetry breaking
in higher dimensions allows for a successful phenomenology
of supersymmetry breaking masses.
One can generate a realistic spectrum, potentially solve the flavor
problem, and address other problems that plague
alternative mechanisms of supersymmetry breaking.
We identify a mechanism for communicating supersymmetry
breaking through the super-Weyl anomaly \cite{sweyl1} \cite{sweyl2} 
\cite{sweyl} which is always present
in hidden sector models in which
supersymmetry breaking is communicated via supergravity. This
is our chief result from which many phenomenological
conclusions follow. We 
also show how separating the supersymmetry breaking sector
introduces a physical cutoff on the theory (below the Planck scale)
and therefore permits perturbative control of supergravity
loops relevant to supersymmetry breaking.

It is important to keep in mind that the framework we develop
in this paper can be applied to any specific theory of higher dimensional
supersymmetry breaking. 
We will identify the important assumptions
for such a theory to agree with our framework.

Our work may have applications to the M-theory scenario of Horava and
Witten \cite{hw} \cite{horava}, which is partly motivated by 
 the possible unification of
gauge and gravitational couplings in higher dimensions. 
More work on this scenario appears in Refs. [9, 15-24]. It should also
apply to string theory models employing D-brane constructions \cite{k}
\cite{lykken}. In addition,
there have been several papers devoted to extra dimensions with
compactification scales of order the weak scale or smaller [27-41]. The
present paper is unrelated to   these proposals; as we
will see our compactification scale is roughly of order the GUT scale.

The outline of the paper is as follows. In 
Section 2,  we review 
the requirements of a successful theory
of supersymmetry breaking.
We discuss existing proposals, and the advantages of the scenario we
present.
In  Section 3, we outline in more detail the higher
dimensional framework we will consider and
contrast it to existing suggestions for breaking
supersymmetry in higher dimensions. In Section 4, we review the
particulars of the 
supergravity  formalism which we will need. We identify
the distinguishing features of a theory derived from higher
dimensions according to our proposal. In Section 5, we discuss
the flavor problem in more detail. We demonstrate why it is endemic
to standard ``hidden'' sector models, but not to ``sequestered'' sector 
theories. In Section 6, we present the heart of our paper, in which
we explain why supersymmetry breaking is necessarily communicated
to the superpartners (including gauginos) at the one-loop level
via the super-Weyl anomaly. In Section 7, we discuss additional
contributions to soft  masses from bulk modes. In Sections 8 and 9, we show
that the mu problem and the supersymmetric CP problem can also be solved.
In Section 10, we discuss the predictions of our scenario. One
of the exciting aspects of our proposal is that it is very
constrained, and therefore very predictive.
We conclude in  Section 11. 

After completing this work we became aware of Ref. \cite{markus} which 
overlaps with some 
of our results in Section 6. 

\section{Supersymmetry Breaking: Constraints and Goals}

Given that supersymmetric partners are yet to be discovered,
it is remarkable how constrained a theory of supersymmetry breaking
is if it is to naturally explain observed features of the
low-energy world. In this section, we enumerate the desired
features of a theory of supersymmetry breaking. We briefly
review how the two standard scenarios, hidden sector supersymmetry
breaking and gauge-mediated supersymmetry breaking fare
under these requirements. 
We then
discuss in a qualitative manner why the sequestered supersymmetry
breaking scenario we have in mind has the potential
to meet the necessary standards. In the remainder of the
paper, we will discuss these issues in greater detail in the
context of the higher dimensional theories.

\begin{enumerate}
\item The first requirement of any successful supersymmetry
breaking theory is to give the correct masses to the superpartners.
Given known constraints (and ignoring experimental loopholes 
\cite{farrar} \cite{farrar}, 
we require masses to be above about 100 GeV. On the 
other hand,
given naturalness constaints, one does not want the 
Higgs, stop, or gaugino masses to be far in excess of a TeV
if the theory is to be reasonably natural. Roughly speaking, this
means that at least some scalar masses and gaugino masses
must be of the same order of magnitude; for example, 
one should not be loop-suppressed
 with respect to the other. Another nontrivial requirement
of a model is that the scalar mass squared for
the superpartners turns out positive.
\item The $\mu$ parameter, multiplying $H_u H_d$ in the superpotential,
should also be between about 100 GeV and  a TeV. The
lower bound comes from the chargino mass, whereas
the upper bound is a naturalness constaint.  Meeting this
requirement is 
a challenge since the $\mu$ parameter appears in a supersymmetric
term; one would like a way
to connect its magnitude to that
of supersymmetry breaking.  
The parameter $\mu B$, multiplying the supersymmetry
breaking scalar mass term $H_u H_d$ should not be substantially
larger than $\mu^2$, although it can be smaller.
\item Flavor changing neutral current constraints provide some
of the most severe restrictions on the supersymmetric spectrum.
The most stringent constraint (if there is no additional CP violation
from the supersymmetry breaking sector) is for the squarks of the first
two generations. In general the constraint depends on several
 parameters. For the case where the gluinos are roughly as heavy as the
squarks the bound is given by \cite{masiero},  
\begin{equation}
\label{flavor1}
\delta_{sq}. (\frac{\rm TeV}{m_{sq}}) < 6 \times 10^{-3},
\end{equation}
where $\delta_{sq}$ denotes the fractional flavor-violation in the
squark mass-squared matrix, and $m_{sq}$ denotes the average squark
mass. 
The constraints from lepton flavor changing events are also quite
severe, in particular coming from the experimental limit on $\mu \to e \gamma$.
Again the exact bound is parameter dependent. 
For tan$\beta \sim {\cal O}(1)$, and a photino with roughly half the
slepton mass, the bound is \cite{masiero}, 
\begin{equation}
\label{flavor2}
\delta_{sl}. (\frac{ 300 {\rm GeV}}{m_{sl}})^2 < 4 \times 10^{-2},
\end{equation}
where $\delta_{sl}$ is the fractional flavor violation in the slepton
mass-squared matrix, and $m_{sl}$ is the average slepton mass. For large
tan$\beta$ there is another important contribution to $\mu \to e
\gamma$. For tan$\beta \sim 50$, a mu-term of $500$ GeV, and 
a photino with roughly half the slepton mass,  the bound is given by,  
\begin{equation}
\label{flavor3}
\delta_{sl}. (\frac{ 300 {\rm GeV}}{m_{sl}})^3 < 4 \times 10^{-5}.
\end{equation}

Of course  the infrared squark and slepton mass matrices  are  to be 
determined by
starting with the mass-matrix obtained at the scale at which flavor and
flavor violation originates and then running it down to low energies using the 
MSSM
renormalization group. Why
the ultra-violet squark mass matrix and running effects should be such
as to produce the extreme level of flavor degeneracy demanded
phenomenologically by eqs. (\ref{flavor1},  \ref{flavor2},
\ref{flavor3}) is the Supersymmetric Flavor
Problem. Its resolution is critical for any candidate realistic theory of
supersymmetry breaking and electroweak symmetry breaking. 
\item The phases of the $A$ and $B$ parameters are constrained
to be small for consistency with the electric dipole moment of the neutron.
That is, CP should be approximately conserved.
\item A less technical requirement for a credible theory (barring
experimental verification) is simplicity.
\item A desirable (though not essential) feature is testability.
\end{enumerate}

Hidden Sector models are defined by the fact that the only couplings
between the supersymmetry breaking sector and the standard sector are
in the Kahler potential and 
are suppressed by $1/M_{Pl}$.

\begin{enumerate}
\item The chief success of hidden sector models is that
it is straightforward to generate positive scalar mass squared.
It is less straightforward to generate the gaugino mass; as
emphasized in Ref. \cite{kaplan}, a singlet VEV which breaks $R$ symmetry
and supersymmetry
is required. This can be a nontrivial requirement in a dynamical
supersymmetry breaking model. It is also a nontrivial requirement
for the weakly coupled heterotic string, where the Green-Schwarz
anomaly induced coupling of the moduli field to the gauginos
is too small compared to the scalar mass \cite{brignole} \cite{gc3}. 
(Even when the
scalar mass is through the anomaly, it is the mass-squared rather
than the mass which is suppressed, thereby enhancing the scalar
relative to the gaugino mass).
\item The $\mu$ problem can be solved through the Giudice-Masiero
mechanism \cite{gm}. That is there can be a term in the Kahler
potential, $\int d^4\theta \Sigma H_1 H_2/M_{Pl}$, where $\Sigma$
is  a singlet which breaks supersymmetry. 
\item CP is generally a problem, without additional symmetries. (See 
  Ref. \cite{ratnir} and references therein.)
\item 
When the above problems are not solved, the theory appears simple;
however attempts to address these inadequacies necessarily
lead to more complicated models.
\item Until we know how the problems are solved, it is difficult
to definitely determine the spectrum. However, to the extent
that the gauge couplings are unified, one can predict the gaugino
mass parameters  corresponding to SU(3), SU(2), 
and U(1)  to be in the ratio of gauge coupling squared.
\end{enumerate}

More recently gauge-mediated models have received a good deal of
attention. (See Ref. \cite{gaugerev} for a review and references.)
In these models, supersymmetry breaking is ultimately communicated
through gauge couplings. The chief objective of these models  was to address
the flavor problem. 
\begin{enumerate} 
\item The success of the spectrum is model-dependent, the greatest
danger being negative mass squared. However, many successful
models were constructed with a desirable spectrum. One of the
very nice features of gauge-mediation is that it is automatic
that gaugino masses arise at one loop, while scalar mass {\it squared}
arise at two, so that scalar and gaugino masses are competitive.
\item There is no compelling solution to the $\mu$ problem,
though there are nice suggestions \cite{mu1} \cite{mu2} \cite{mu3}
\cite{mu4}. What makes
it so difficult is that even if one accepts a small parameter
to suppress $\mu$ in the superpotential, the same small parameter naturally
multiples the mass {\it squared} parameter $\mu B$, in which case the $\mu$
term is too small compared to $\mu B$ \cite{mu1}.
\item The chief success of gauge-mediated models is
that they have no flavor problem.
\item They generally have a CP problem.
\item They are often very complicated. Particularly when
one incorporates the requirement of a $\mu$ parameter, there
is no model which seems a likely candidate for describing the
real world.
\item The models are predictive.  They give the same
prediction for the ratio of gaugino mass as the hidden
sector models. However, the spectrum of scalar masses
is determined by gauge couplings and charges and is therefore
distinctive.
\end{enumerate}

We now introduce the idea of a ``sequestered'' sector, which
is basically a hidden sector that is truly hidden. Supersymmetry
breaking can occur anywhere except on the 3-brane where the standard
model particles are confined. Gravity in the bulk will permit
the communication of supersymmetry breaking. If the 
``compactification'' scale, determined by the separation of the
supersymmetry breaking  from the visible sector 3-brane is sufficiently small
(that is the separation exceeds the Planck length), the theory
is well under control. There are no direct couplings between the sequestered
sector and the standard sector above the compactification scale.
All gravity communication can be addressed perturbatively,
since gravity mediated loops are suppressed by the ratio of
the compactification to Planck scale.
There can however be direct couplings between
the sectors, depending on what
matter resides in the bulk. We will discuss later the possibilities
both with and without bulk matter which couples to both
sectors (aside from gravity).
\begin{enumerate}
\item One can obtain a successful spectrum. The key ingredient
is the as yet neglected effects of the super-Weyl anomaly, which always
introduces a coupling of the auxiliary field of the gravitational
multiplet to both the gaugino and scalar masses. The
auxiliary field obtains a vacuum expectation value
through its coupling to the supersymmetry breaking
sector. It couples in turn to the standard
model fields because of the breaking of scale
invariance. The couplings are determined through
the {\it supersymmetric} renormalization group functions. 
 In a usual hidden sector scenario, these couplings are present
but are suppressed relative to the leading term. We assume
the bulk matter is such that these are the {\it dominant} contributions
to the gaugino and squark mass squared (the sleptons are lighter
and have competitive additional contributions). One satisfying
aspect of this coupling is that there is no one-loop contribution
to the scalar mass squared; with this ``anomaly-mediation''
the gaugino and scalar mass (for fields with the appropriate
gauge charge) are comparable. We find however
that for fields which transform only under nonasymptotically
free gauge theories, the mass squared due to the anomaly contribution
is negative. This implies the necessity of additional mass contributions
to the scalar mass squared. These can arise from many sources that
will be discussed in Section 7. 
\item The $\mu$ parameter can be generated without generating an excessively 
large $\mu B$
parameter due to the constraints from the coupling of the gravitational
multiplet.
\item CP suppression is straightforward. Even without any further
analysis, it is straightforward to  have CP violated on our 3-brane,
but nowhere else.  We then find 
the absence of new phases in the  $A$ and $B$ terms.
This implies  a natural solution of the SUSY strong CP problem.
\item Flavor violation is automatically suppressed by the dominant
anomaly-mediated contribution to the squark mass squared. 
Additional scalar mass contributions may or may not be flavor preserving.
It is straightforward to make models in which flavor preservation
is automatic.
\item The ``model'' is straightforward. In fact there
is little model building required. Most of the results
we present in this paper are straightforward consequences
of our fundamental assumption that we derive supersymmetry-breaking
from a higher dimensional theory.
\item Sequestered supersymmetry breaking
 is very predictive. The ratio of gaugino masses
depends on the beta functions, rather than simply the gauge
coupling as for the other two scenarios.
There is a nearly degenerate wino/zino LSP, of
which the zino is the lighter.  We predict $A$-terms proportional to the 
corresponding 
Yukawa couplings.
\end{enumerate}

\section{The Framework}

The scenario we consider involves supersymmetry breaking in a higher
dimensional theory. The first question this brings up is
where are the standard model fields? We have implicitly assumed
until this point that the standard model fields are confined
to a four dimensional boundary. This is the only consistent
assumption if we are to avoid rapid blow-up of their couplings in the
higher dimensional theory and a very low string scale. 
The basic problem is that if one
assumes a coupling of order $M^{d-4}$ in a compactified $d$-dimensional
theory, where $M$ is some fundamental scale such as the $d$-dimensional
Planck scale, 
it would induce a dimensionless coupling of order $(r_c M)^{d-4}$ in the
effective four-dimensional theory,
where $r_c$ is the compactification radius. 
When the compactification
scale is well  below $M$,  this would imply
a small four-dimensional coupling.
However, we know that several of the standard model couplings are of
order unity. Therefore, 
there cannot be a large separation between the compactification scale
and the scale $M$  when there are standard model fields in the bulk.
In this case, one runs the risk of uncontrolled  flavor violation,
since there is no large scale separation. So in this paper
we choose to examine a scenario in which the standard model fields
are confined to a four-dimensional subspace, or ``3-brane'' in the higher
dimensions. This is the ``visible sector''.

The next question is what we assume about the ratio of the
higher dimensional compactification radii. In fact, our
results can be examined without explicitly assuming any
particular scheme. There might be six large dimensions
(recall here large means within two orders of magnitude of the Planck
scale), there might be one, or various other possibilities in between.

A further question is what we  assume to be in the bulk higher
dimensional spacetime. We will first consider
the effects of the gravitational multiplet, and
then consider the effects of additional bulk matter fields.
It is very important that we assume there
are no bulk fields that have large flavor-dependent
couplings to matter fields. What large means
will be made clear later on.. This is nonetheless
a strong assumption. For example, in the Horava-Witten
setup, it would mean assuming the dilaton is heavy. It is hard to assess
the likelihood of this assumption being correct, since we simply do not yet
know how moduli states obtain a potential. Presumably, the
states do have a potential and are massive with a well-defined
minimum. Our assumption is that this mass is above the compactification
scale. It could be that these states obtain mass only after
superysymmetry breaking (though cosmology argues against that \cite{kaplan}).
In that case our assumption would simply not be true.

There are then two scenarios for supersymmetry breaking in the
bulk which we have in mind when
setting up our framework.  Supersymmetry breaking can occur
on another 3-brane via  a truly
hidden sector. Alternatively, supersymmetry breaking can occur
through a nonlocal effect in the bulk. That is the conditions
for supersymmetry can be different in different slices of the bulk.
There are several examples where this is the case, including
Scherk-Schwarz \cite{ss}, and rotated branes \cite{rot1} \cite{rot2}
\cite{rot3}. From
the point of view of the low-energy theory below the compactification
scale, these theories will look very similar. There will be an
effective supersymmetry breaking sector whose fields do 
not have direct couplings
to those of the standard model fields. The theories will differ
in the precise field responsible for supersymmetry breaking,
the precise form of the low-energy potential for this field,
and in the relation between fundamental scales of the high dimensional
theory and the supersymmetry breaking scale observed in the visible sector.
These details are not relevant to the
results we find in the four-dimensional effective field
theory analysis, which
would apply to any of these models. 
It is nonetheless of interest to explicitly
analyze the high-energy theories
to see how well they accomodate the assumptions we
outline later.
A more detailed discussion of the non-local
supersymmetry breaking scenarios from the four-dimensional
effective field theory will be presented in a subsequent paper \cite{rs}.

A final example of a theory involving supersymmetry breaking in
higher dimensions is  that of Horava and Witten \cite{hw, horava}. 
It is interesting
in that in the presence of the full dilaton multiplet,
supersymmetry breaking is nonlocal; one can find a supersymmetric
solution on each wall. However, if the dilaton multiplet is heavy,
supersymmetry breaking could be broken dynamically  on the invisible
boundary, making it a truly hidden sector.

There have already been analyses of the low-energy effective
field theory for some models derived from string/M-theory. Ref. \cite{gc2}
evaluated the loop-induced 
contribution to the scalar and gaugino masses. Whereas the former
were of order $m_{3/2}^2/M_{Pl}$, the latter were much smaller
of order $m_{3/2}^3/M_{Pl}^2$. This is the standard small
gaugino mass problem \cite{gaugino}. 
Ref. \cite{gc3} also included the ``dilaton''.
This is important because the direct coupling
of the large compactification modulus (with the supersymmetry
breaking auxiliary component) generates
at tree level neither scalar nor gaugino mass.
However,   through mixing between the dilaton and the compactification
modulus, one generates a  coupling of order $\alpha$ for both the
gaugino mass and the scalar mass {\it squared}. Here $\alpha$ 
is a parameter which determines the effective vacuum energy
on the boundaries and  bulk, and arises from the internal
components (assuming Calabi Yau compactification of six of the dimensions)
of the three-form tensor necessary to stabilize the configuration.
In the weakly coupled heterotic string, where $\alpha$ arises
from the Green-Schwarz mechanism, the gaugino mass is too small.
In the M-theory context, Ref. \cite{gc3} 
requires that the gauge and gravity couplings are unified,
which gives $\alpha$ of order unity, permitting a phenomenologically
acceptable ratio of the gaugino and scalar masses. However, if indeed
this is the dominant contribution to the scalar mass squared,
it is difficult to imagine a resolution of the flavor problem.
In general, arbitrary renormalization effects at the Planck
scale would generate flavor-dependent couplings for the dilaton
(or other shape moduli which would also mix), implying nondegenerate
scalar mass squared.

In this paper, we will show there is generally an additional anomaly-induced
contribution to both the scalar and gaugino mass squared. We assume
that the content and scales of the theory are such that this is
the dominant squark mass contribution. For most  of the
paper we will assume that this is because flavor dependent contributions
mediated by light bulk states are simply absent. 
We will further discuss the significance of the flavor constraints in 
Section 10.

It is important to recognize that the framework we establish
here can be applied to any specific model. For any
given example, one would need to establish if the assumptions
we find necessary apply. In a subsequent paper, \cite{rs}
we will apply these considerations to ``nonlocal'' theories
of supersymmetry breaking.

\section{The Four-Dimensional Effective Theory} 

In order to derive the low-energy consequences of the existence
of higher dimensions, one could take several approaches.
One could start from a fundamental theory and derive in the
higher dimensions the exact mechanism for communicating
supersymmetry breaking via auxiliary components in the higher dimensional
theory. This approach was taken in a toy model in Ref. \cite{peskin}.
Alternatively, one can assume that supersymmetry breaking is indeed
communicated and derive the effective low-energy theory consistent with
this and other facts we know about a specific higher dimensional theory,
while accounting for general covariance of the low-energy theory.
In this section we follow the latter approach and derive
  the special form of the effective {\it
  four-dimensional} 
supergravity Lagrangian below the compactification scale, $\mu_c$,
corresponding to the higher-dimensional scenario in which the visible
 and hidden sectors live on separate 3-branes. We begin by
reviewing the general form of four-dimensional supergravity coupled to
matter, 
as detailed for example in Ref. \cite{wessbagger}. This will allow us to
 point out the important features and to set notation.

\subsection{General Four-Dimensional $N=1$ Supergravity}

There are several formulations of supergravity; one can 
write the Lagrangian explicitly in terms of $E$ and ${\cal E}$,
the vielbein superfields  along
the lines of \cite{wessbagger} or one can use a  compensator
formalism, along the lines of Refs. \cite{1001} \cite{sweyl}.
 We choose to use the Wess and Bagger formalism
but we explicitly isolate the  complex spin-0 auxiliary field which
can acquire a supersymmetry-violating but Lorentz invariant
expectation value. It will be convenient for us to formally define a
flat-space chiral
superfield, $\Phi$, to house this auxiliary field, 
\begin{equation}
\Phi \equiv 1 + F_{\Phi} \theta^2.
\end{equation}
We stress that $\Phi$ is not a separate chiral superfield in curved
superspace, but rather just a formal device for separating out the
scalar auxiliary field of the off-shell supergravity multiplet. This
device will be very convenient later when we consider supersymmetry
breaking in the the flat space limit.
In terms of  $\Phi$ we have,   
\begin{equation}
{\cal E} =e \Phi^3 + ...
\end{equation}
where $e$ is the determinant of the vielbein and the ellipsis contains
supergravity fields of non-zero spin. 
This same field $\Phi$ appears in the kinetic terms since
we can rewrite \cite{us,1001} (again dropping fields with non-zero
spin), 
\begin{equation}
E \propto \Phi \Phi^\dagger.
\end{equation}
For this reason, the coupling of the field $\Phi$ is determined
by that of the metric determinant,
 and  it will enter in a way precisiely
determined by the conformal scaling of any operator in the Lagrangian. 

The general Lagrangian (up to two-derivative order
in the low-energy expansion) for supergravity coupled to matter can then be
written,
\begin{eqnarray}
\label{lag}
{\cal L} &=& \sqrt{-g}\{\int d^4 \theta f(Q^\dagger, e^{-V} Q)
\Phi^\dagger \Phi ~+~ \int d^2 \theta (\Phi^3 W(Q) +
\tau(Q) {\cal W}_{\alpha}^2) + {\rm h.c.} \nonumber \\
&-& \frac{1}{6} 
f(\tilde{q}^\dagger, \tilde{q})(R + 
{\rm vector~ auxiliary~ terms}  + {\rm gravitino~terms}\},
\end{eqnarray}
where we have employed flat-superspace notation in the first line, 
with the understanding that derivatives of the matter fields are to be 
made covariant  with respect to
gravity and to the auxiliary gravitational vector field.
We denote the matter chiral superfields by $Q$, their lowest bosonic
components by $\tilde{q}$, vector superfields by $V$, and their
supersymmetric field strengths by ${\cal W}_{\alpha}$. We have dropped
color and flavor
indices for notational simplicity. $R$ is the spacetime curvature scalar.
The
function $f$ can be written in the form,
\begin{equation}
\label{Kdef}
f \equiv - 3 M_{Pl}^2 e^{- K/3 M_{Pl}^2},
\end{equation}
where $M_{Pl}$ denotes the four-dimensional reduced Planck mass, and
where $K$ is defined to be the supergravity Kahler potential. 
With this definition of $K$, a canonical $K=q q^\dagger$ leads
to  normalized fields with no higher dimensional two derivative terms,
as shown  below in Eq. \ref{canonical}.

The supergravity lagrangian eq. (\ref{lag}) 
is not in the canonical (and perhaps more familiar) form where the
Einstein action has a field-independent coefficient of 
$M_{Pl}^2/2$.  To obtain the canonical form we have 
to eliminate the field-dependence 
 by redefining the  metric by a Weyl transformation, 
\begin{equation}
\label{weyl} g_{\mu \nu} \rightarrow   e^{K/3 M_{Pl}^2}
 g_{\mu \nu}. 
\end{equation}
This transformation then changes the $K$-dependence (or $f$-dependence) 
of the remaining terms in the lagrangian.  
After  Weyl transforming and integrating out  auxiliary
fields, the resulting 
Lagrangian  is
\begin{eqnarray}
\label{canonical}
{\cal L}  &=& 
\sqrt{-g}\{\frac{M_{Pl}^2}{2} R + K_{i j}(\tilde{q}^\dagger,
\tilde{q}) D_{\mu} \tilde{q}^{i \dagger} D^{\mu} \tilde{q}^j - 
{\cal V}(\tilde{q}^\dagger, \tilde{q}) \nonumber \\
&-& \tau(\tilde{q}) (F_{\mu \nu} F^{\mu \nu} + i F_{\mu \nu} \tilde{F}^{\mu
  \nu}) + {\rm h.c.} 
+ {\rm fermion~terms},
\end{eqnarray}
where the Kahler metric is given by,
\begin{eqnarray}
K_{ij}(\tilde{q}^\dagger, \tilde{q}) &\equiv& 
\frac{\partial}{\partial \tilde{q}^{i \dagger}} \frac{\partial}{\partial 
\tilde{q}^j} K, 
\end{eqnarray}
and the scalar potential is given by
\begin{equation}
{\cal V} = e^{K/M_{Pl}^2} \{(\frac{\partial W}{\partial \tilde{q}^i} + 
\frac{W}{M_{Pl}^2}
\frac{\partial K}{\partial \tilde{q}^i}) K^{-1~ i j} 
(\frac{\partial W^\dagger}{\partial \tilde{q}^{j \dagger}} + 
\frac{W^\dagger}{M_{Pl}^2}
\frac{\partial K}{\partial \tilde{q}^{j \dagger}}) ~-~ 3 
\frac{|W|^2}{M_{Pl}^2}\} + \frac{g^2}{2}(\frac{\partial K}{\partial
  \tilde{q}} t^a \tilde{q})^2,  
\end{equation}
where the last term is the gauge D-term scalar potential.
This is the more familiar form of supergravity lagrangian. However, this
form obscures many of the features we are interested in, and
we will therefore mostly employ eq. (\ref{lag}).

In this formalism the field-dependent gravitino mass is given by, 
\begin{equation}
\label{gravitino}
m_{3/2} = e^{K/2 M_{Pl}^2}\frac{|W|}{M_{Pl}^2}.
\end{equation}
This becomes the physical gravitino mass when we replace all fields by
their vacuum expectations and after tuning the cosmological constant to
zero.

It is also edifying to observe the origin of the
scalar mass squared in Eq. (\ref{canonical}).
When supersymmetry is broken, there are two things which
carry this information; the VEV of the field in the goldstino mutiplet
$\Sigma$ and the auxiliary component of the $\Phi$ field (which translates
into dependence on $W$ in Eq. (\ref{canonical})).  From 
Eq. (\ref{lag}),  one can verify explicitly or by rescaling
that there is no tree-level contribution to the $\tilde{q}$ scalar
mass from $F_{\Phi}$. The source of mass can then only be
terms which couple $\Sigma$ and $Q$ directly in $f$ or from the
terms proportional to $R$. Clearly in a flat-space background
in which the cosmological constant has been cancelled, the second term
does not contribute. So the {\it only} source of tree-level scalar
mass is the ``curvature'' terms in $f$ which introduce {\it direct}
couplings between the so-called hidden sector and the visible sector.
This fact is obscured somewhat in Eq. (\ref{canonical}), but is nonetheless
true. It is manifested by the fact that when $W$ is chosen to cancel
the cosmological constant, the source of the scalar mass squared
is $\partial W/\partial \Sigma$, and not $W$.

\subsection{Effective Supergravity in the 3-Brane Scenario}

Let us now specialize to  the  effective theory describing
physics of an initially higher dimensional theory  below the
compactification scale, $\mu_c < M_{Pl}$ in the 3-brane scenario. 
For simplicity, in this section, we neglect bulk fields
other than four-dimensional supergravity. 

In this case, 
 the effective
four-dimensional theory consists of supergravity coupled to the visible
and sequestered sector fields.\footnote{Exactly how the higher dimensional
theory reduces to four-dimensional supergravity is  an
interesting and subtle issue.  For a detailed explicit discussion of 
the dimensional reduction in the (simpler) case where the bulk fields are
gauge fields rather than supergravity, see ref. \cite{peskin}.}
The essential point in deriving the constraints imposed from
the above assumptions is that there is no direct coupling in the
higher dimensional theory between the sequestered sector fields and
the visible sector fields.  We are quite literally hiding the hidden
sector. 
 This decoupling means that there are no allowed operators
(at the level of $1/M_{Pl}^2$) at tree level with direct couplings
between sequestered and visible sector fields.
 As we will see, this is the key to the resolution of the
flavor problem.
Nevertheless, there will be couplings generated at the radiative
level between the two sectors. That will be the subject of the 
sections that follow.

The above observation gives  a  powerful constraint on the form of the
four-dimensional supergravity Lagrangian effective below $\mu_c$, namely
that  if the 
four-dimensional supergravity fields are formally switched off,
\begin{equation}
g_{\mu \nu} = \eta_{\mu \nu}, ~~ \Phi = 1, 
\end{equation}
then the visible and hidden sectors should decouple. It follows that 
Eq. (\ref{lag}) must take the special form 
\begin{eqnarray}
\label{special}
f &=& -3 M_{Pl}^2+ f_{vis} + f_{hid}  \nonumber \\
W &=& W_{vis} + W_{hid} \nonumber \\
\tau {\cal W}^2 &=& \tau_{vis} {\cal W}_{vis}^2 + \tau_{hid} {\cal
  W}_{hid}^2,
\end{eqnarray}
where the functions $f_{vis}, W_{vis}, \tau_{vis}$ depend
only visible sector fields, and $f_{hid}, W_{hid}, \tau_{hid}$  are
functions of only hidden sector fields. 

As can be seen, the fact that the visible and hidden sectors are
physically separated in the underlying higher dimensional theory is
reflected very simply in the special form of Eq. (\ref{lag}) given
above. On the other hand, the Kahler potential defined by
eq. (\ref{Kdef}) is given by
\begin{equation}
K = - 3 M_{Pl}^2 ~{\rm ln}(1- \frac{f_{vis}}{3 M_{Pl}^2}
- \frac{f_{hid}}{3 M_{Pl}^2}) \nonumber \\
\end{equation}
and is not simply additive for the two sectors. 
The induced couplings between the two sectors are consequences
of the fact that both sectors couple to a common gravitational
muliplet and because of the Weyl rescaling done
to obtain a canonical $M_{Pl}$. It is important
to recognize that the separation is manifest in $f$, and not $K$.

One can use this Kahler potential to derive masses in the low-energy
theory; however, it clearly obscures the fact that the theory
originated in two decoupled sectors. Of course, in principle
a four-dimensional theory could have a Kahler potential of this
very special form; however, there is no symmetry which would
maintain it in the presence of radiative corrections.
Here this separation is nonetheless ``natural'' in that
it is enforced by the geometry.
It is clearly simpler to
work directly with $f$ since the Weyl-rescaling given by
Eq. (\ref{weyl}) to obtain the canonical form of Eq. (\ref{canonical})
 completely obscures the simplicity of Eq. (\ref{special})
by introducing many apparent interactions between the two sectors. Many
of the special properties of our scenario then appear as the result of
``miraculous'' cancellations. We will therefore work with  the
non-canonical form of the lagrangian, Eq. (\ref{lag}), where the special
properties are manifest. 

All the subsequent conclusions follow from the above special form of the
gauge kinetic function, superpotential, and Kahler potential and therefore
can be understood solely in four-dimensional language.  In fact, one
can ask whether we really exploit the higher dimensions. The answer
is we do not see any way to guarantee the above form without such an 
assumption.
It is clearly not protected by symmetries of the four-dimensional world.
Without the assumption of a deep underlying reason for
separating the elements of the Lagrangian in this way, it would
be an ad hoc assumption not stable to radiative corrections (at
least for the Kahler potential).

For the purposes of the rest of the paper, we will take
$f_{hid}$ and $W_{hid}$ 
to contain only a chiral mutiplet $\Sigma$ in which the Goldstino
resides as the fermionic component.   
It will become clear that our general conclusions rely on the
assumption of the factorization of $K$ and not on the specific
form of $f_{hid}$. It is therefore simplest to take the minimal sequestered
sector and to ignore any other potential light fields which might
reside there.  The superfield $\Sigma$ may be  composite or fundamental.

We assume that both the lowest component and auxiliary components
of $\Sigma$ can acquire nonvanishing expectation values.
\begin{equation}
\langle \Sigma \rangle =\langle \sigma \rangle + \Lambda_H^2 \theta^2.
\end{equation}
Here $\Lambda_H$ is the scale of supersymmetry breaking 
and $\langle \sigma \rangle$ is assumed to be well below
the Planck scale (otherwise we would work in terms of redefined fields).
The VEV of the lowest component $\langle \sigma \rangle$ will not play
an important role in any case. 

\section{The Supersymmetric Flavor Problem}

In the previous section, we remarked that the squark mass squared
must arise (classically) from direct couplings between the so-called hidden
and visible sectors. This clearly permits the possibility
of dangerous flavor changing couplings due to dimension
four operators in the Kahler potential which couple hidden
sector and visible sector fields.  In fact, from a field
theory perspective, we know these flavor violating couplings
are present. This is because radiative corrections
in the MSSM require Yukawa dependent counterterms, due
to the logarithmically divergent running.

This also makes it clear why higher dimensional supersymmetry
breaking has the potential to eliminate dangerous flavor violation.
First of all, the sectors are decoupled at tree-level. Second,
the addition of higher dimensions and the consequent separation
between the sequestered and visible sectors actually improves
convergence in the effective theory. This is clear
because {\it there is no counterterm involving hidden and
visible sector fields.} Any coupling between the two sectors
arises from a finite supergravity calculation. In fact,
the situation is even better because the introduction of a cutoff
scale (the compactification scale) beneath the Planck scale
means that higher loops are further suppressed by powers of $\mu_c/M_{Pl}$.

Let us see how the flavor problem manifests itself in the context of purely
four-dimensional hidden sector models of supersymmetry breaking. The
general supergravity Lagrangian in four dimensions will have
Planck-suppressed interactions between the hidden and visible sectors
given
by 
\begin{eqnarray}
f &=& (1 - \frac{h}{M_{Pl}^2} \Sigma^\dagger \Sigma) Q^\dagger e^{-V}
Q + ... \nonumber \\
\tau &=& (\frac{1}{g^2} + \frac{k}{M_{Pl}} \Sigma) + ...,
\end{eqnarray}
where $h, k$ are order one dimensionless constants.
Even if one were to posit that these interactions were absent at tree
level, 
they would certainly be induced at loop level at the Planck
scale. There need be no suppression by a small dimensionless parameter
because gravity is strong at the Planck scale. On the other hand it is
consistent to assume that there is no interaction between the visible
and hidden sectors in the superpotential, 
\begin{equation} 
W = W_{vis}(Q) + W_{hid}(\Sigma),
\end{equation}
since this relation is radiatively stable due to the non-renormalization
theorem. After putting in the hidden sector VEV, we find soft visible
sector masses, 
\begin{eqnarray}
m_{\tilde{q}}^2 &=& h \frac{\Lambda_H^4}{M_{Pl}^2} \nonumber \\
m_{gaugino} &=& k \frac{\Lambda_H^2}{M_{Pl}}.
\end{eqnarray}
If $\Lambda_H \sim \sqrt{v ~M_{Pl}}$, where $v$ denotes the weak scale, 
we obtain weak-scale soft masses as
desired. 

The flavor problem resides in the fact  that there is no reason for the
$h$ couplings of the squarks to the hidden sector, and consequently the
ultraviolet squark masses, to respect flavor. It is very difficult to
understand why the string or Planck scale physics (implicitly integrated
out to yield the effective supergravity lagrangian) should very precisely
respect flavor when the lower-energy physics certainly does not. In
early papers the flavor problem was obscured by working in terms of the
Kahler potential, $K$,  instead of $f$, and assuming, 
\begin{equation}
\label{separateK}
K = |\Sigma|^2 + Q^\dagger e^{-V} Q.
\end{equation}
The proposed justification was that this form of Kahler potential leads to
the minimal renormalizable kinetic terms in
Eq. (\ref{canonical}). Formally, it appears that there are no
direct interactions between the hidden and visible sectors in both the 
 Kahler
potential and superpotential, and therefore any visible soft terms that
result must be purely gravity-mediated. Since gravitational couplings
are  flavor-blind it was expected that the soft terms should be as
well. This argument is quite false (as is now generally
appreciated). Using Eq. (\ref{Kdef}) we find that the proposed Kahler
potential  corresponds to
\begin{equation}
f = - 3 M_{Pl}^2 + |\Sigma|^2 + (1 + \frac{|\Sigma|^2}{M_{Pl}^2}) 
\overline{Q} e^{-V} Q + ...
\end{equation}
which is flavor-blind. However, we see that the soft masses arise out of
a direct, albeit Planck-suppressed, coupling between the visible and
hidden sector fields, not as the result of a supergravitational
exchange. The real source of this coupling is that we have implicitly
integrated out Planck-scale (string) states which may couple to both the
visible and hidden sectors. There is absolutely no reason to believe
that their couplings are flavor-blind;  on the contrary one would expect
at least some of them to distinguish the flavors \cite{lounir}. 
The second point is
that it is absurd to expect renormalizable matter kinetic terms {\it
  after Weyl rescaling} in a
non-renormalizable supergravity theory. Therefore the special form
of eq. (\ref{separateK}) is highly unnatural and the flavor problem
prevails in the context of four-dimensional hidden sector models. 

Let us now turn to our 3-brane scenario. From Eq. (\ref{special}) it
is clear that there is no direct coupling between the hidden and visible
sectors, and that gravity really is the only intermediary. Of course just as in
purely four-dimensional scenarios, we have implicitly integrated out
any Planck-mass bulk states in getting to the higher dimensional effective
field theory. One expects  some of these states to couple to visible
sector matter in a flavor-dependent way.
 However, in this scenario, the couplings induced by heavy
states are exponentially suppressed.
This can be more precisely understood by
thinking of any such potential effects as due to the exchange of one or
more bulk propagators. Since the 3-branes break translation invariance
in the extra dimensions, it is useful to  consider 
a bulk propagator in position space. It must extend from the
visible sector 3-brane out to a point roughly a distance $r_c =
1/\mu_c$ away in the extra dimensions. 
This could be a point on a separate 3-brane if we are
considering supersymmetry breaking localized there, or a
typical point in the bulk in ``non-local'' supersymmetry breaking 
 scenarios. Therefore, a
coordinate space propagator for a bulk state of mass $m$ will be
suppressed by $e^{- m r_c} \sim e^{-m/\mu_c}$. This is one of the
central distinguishing features of the seqestered sector scenario: Bulk
states with masses $m \gg \mu_c$ will have their contributions
to visible sector supersymmetry breaking sharply cut off. If all bulk
states with flavor-dependent couplings to the visible sector satisfy $m
> \mu_c$, then their flavor-dependent supersymmetry breaking effects are
suppressed. In purely four-dimensional theories there is no such ``switch'' for
turning off flavor-dependent supersymmetry breaking effects. 

Until Sections 7 and 8, we will assume that the only bulk states lighter
than $\mu_c$ belong to four-dimensional 
supergravity. Since masses (and off-shell momenta)
larger than $\mu_c$ are irrelevant for communication of supersymmetry-breaking
to the visible sector, we can study that process in the four-dimensional
supergravity theory effective below $\mu_c$. 
The only super-gravitational field that enters into nonderivative terms and can
therefore give rise to soft visible masses is the  chiral
field $\Phi$. Indeed it  generally has a
supersymmetry-breaking $F$-term expectation value, as illustrated 
in  Appendix 1. In general, 
\begin{equation}
F_{\Phi} \sim {\cal O} (F_{\Sigma}/M_{Pl}) = \Lambda_H^2/M_{Pl}.
\end{equation}
It is clear that only through this term can supersymmetry breaking be
communicated to the visible sector. However it is easy to
see that in fact soft visible masses are not generated
classically. First since $\Phi$ does not couple to the gauge field
strength the gaugino masses vanish.

It is clear that the superpotential is irrelevant
to the soft scalar mass squared (in the absence of fundamental
mass terms or visible sector VEVs). The only term which is then
relevant is
\begin{equation}
{\cal L}_{visible 
kinetic} = \int d^4 \theta f(Q, Q^\dagger) \Phi \Phi^\dagger \, .
\end{equation}
It is clear that this term does not generate a soft scalar
mass since $\Phi$ can simply be supersymmetrically redefined away 
by $Q \Phi \rightarrow Q$. This can also be readily found
explicitly by evaluating the $d^4 \theta$ integral which gives
\begin{equation}
{\cal L}_{aux} = f_{vis}(\tilde{q}^\dagger, \tilde{q}) |F_{\Phi}|^2 
+ \frac{\partial^2 f_{vis}}{\partial \tilde{q} \partial
  \tilde{q}^\dagger} |F_Q|^2 +
\frac{\partial f_{vis}}{\partial \tilde{q}} F_Q F_{\Phi}^\dagger  +
3 W_{vis}(\tilde{q})F_{\Phi} + 
{\frac{\partial W_{vis}}{\partial \tilde{q}} F_Q + \rm h.c.} + ...  
\end{equation} 
Integrating out $F_Q$ gives a potential for $\tilde{q}$ of the form, 
\begin{equation}
\label{vispot}
V_{vis}(\tilde{q}) = \frac{| \frac{\partial W_{vis}}{\partial
      \tilde{q}} + \frac{\partial f_{vis}}{\partial \tilde{q}}
\overline{F_{\Phi}}|^2}{\frac{\partial^2 f_{vis}}{\partial \tilde{q} \partial
  \tilde{q}^\dagger}} - 
6 {\rm Re}~ W_{vis}(\tilde{q})F_{\Phi} -f_{vis} |F_\Phi|^2. 
\end{equation}
It is clear that the mass term cancels.
 This may appear to be an unfortunate accident. We will see in
the next section that this is not an accident and that it is
 far from unfortunate. Masses will indeed be generated at the
radiative level, and the leading contribution
to the squark mass matrix will be flavor-symmetric.
Clearly, the absence of tree level terms in the fundamental
Lagrangian which give mass means that the mass generation
in the effective low-energy theory must give a finite 
calculable (or potentially vanishing)  squark mass.

Before proceeding, let us briefly investigate other potential
flavor-violating operators. There could be higher dimension operators
involving purely visible sector fields. Because this are
 suppressed by the Planck scale, they are negligible. Other
possible operators suppressed only by the compactification scale
could be generated by integrating out Kaluza Klein modes of bulk moduli. Again,
these are clearly sufficiently suppressed
if the compactification scale is sufficiently high
which it will be. Finally, flavor
violating masses can and will be generated in loops
in the low-energy theory due to flavor-violating Yukawa couplings.
However, these will be finite and cut-off by the compactification scale.
Flavor violation in the low-energy theory is
 calculable. For the known Yukawa couplings, we are safe.
The problem which plagues ordinary hidden sectors is the unknown
counterterms, which are absent in the higher dimensional  scenario.

\section{Anomaly-Mediated Supersymmetry Breaking}

Although we have argued that there are no tree-level
supersymmetry breaking visible sector masses, 
this is not true at the quantum level. We will
show that certain anomalous rescaling violations provide a previously
unrecognized mechanism for communicating supersymmetry breaking.
It will turn out that ``anomaly-mediated'' supersymmetry
breaking generates a one-loop gaugino mass and degenerate
two-loop scalar mass squareds. We stress that this communication
of supersymmetry breaking is not optional; it operates
within {\it any} hidden sector model.
In most hidden sector models, there are
 larger tree-level contributions, at least
to the scalar mass squared. As we have discussed,
this is not true for sequestered supersymmetry breaking,
so the gaugino and squark masses we derive here
will be the {\it dominant} contribution to the masses. 
 However, as we will see, in order to obtain a
realistic spectrum, comparable bulk loop effects of the type discussed
in Section 7 must also be present.

 We remind the
reader of the super-Weyl anomaly and discuss its implications
for the gaugino mass. We will then extend the analysis to
the effects of rescaling anomalies on the scalar sector,
where we will show that scalar masses are also generated. 

Let us begin with a visible sector consisting of pure super-Yang-Mills
theory. From Eq. (\ref{lag}) it appears that the visible sector
Lagrangian does not couple to $\Phi$. This corresponds to a classical
invariance of the visible sector Lagrangian  under super-Weyl
transformations, where $\Phi$ (or more precisely the determinant of the
super-vielbein) is multiplied by an arbitrary chiral superfield. In
particular this allows us to completely 
Weyl-transform away $\langle \Phi \rangle$ from the supergravity
multiplet which couples to the Yang-Mills theory, thereby obtaining a
canonical supersymmetry preserving gravitational background. If the
classical super-Weyl invariance were exact this would imply that the
visible gauginos are massless. However it is known that in the presence
of the super-Yang-Mills sector, the super-Weyl transformation  is
anomalous, and results in the shift \cite{sweyl1,sweyl2,sweyl}, 
\begin{equation}
\label{taushift}
\tau \to \tau - 2 b_0 {\rm ln}(\Phi)
\end{equation}
where  
$b_0$ is the gauge theory one-loop beta function coefficient reflecting the
anomaly in Weyl (scale) invariance.

Now it is known that the dependence on the leading component
of $\Phi$ goes away; since we have already
set the leading component to 1, this is manifest. However,
there is nontrivial dependence on the auxiliary component of $\Phi$.
Explicitly, after evaluating the $\theta$ integral there
is a gaugino mass term
\begin{equation}
\label{gauginomass}
m_{gaugino}=-b_0 g^2 F_\Phi
\end{equation}
where $g$ is the gauge coupling.

This is a critical result. It means that even when there is no
direct coupling of fields of the supersymmetry breaking sector
to matter, there is nonetheless a gaugino mass which appears
at one loop. This is true quite generally.
It is most important of course when there is no larger term.
In the absence of tree-level couplings, this would be the dominant
gaugino mass contribution. 
The prediction in such models is that gaugino masses appear in the
ratio of their beta functions. 

There are many ways to derive the above anomalous coupling but it can
be thought of as a result of a violation of super-Weyl invariance.
This is associated with the cutoff dependence  of a non-finite
theory and should have effects wherever cutoff effects appear.  
Although the renormalizable physics is independent of 
 the detailed nature of the cutoff, we point out that in the present
scenario there is a physical cutoff provided by the string scale (or
whatever scale marks the onset of new gravitational physics). In fact results 
of the form 
of Eq. (\ref{gauginomass}) were  derived in certain string-based models by 
accounting for string-threshold corrections \cite{lust}. Here we will extend 
these 
results and demonstrate their generality within effective field theory. Soft 
masses 
proportional to beta functions have also been derived in the string-based 
models discussed in Refs. \cite{wu}. 

We now consider the further consequences of the scale anomaly
and  derive the scalar mass squared. For now, we treat the 
 the bulk classically.
 That is, we neglect supergravity loops
as well as those of other fields which might be present in the bulk.
It should be clear (up to $M_{Pl}$ suppressed interactions
induced by Weyl-rescaling) that the hidden sector dynamics
decouples from the visible sector, and can be integrated out.
The only effect of the hidden sector dynamics will be through the $\Phi$
field (that is gravity) which couples to both sectors.
\begin{eqnarray}
\label{renolag}
{\cal L}_{eff}(M) &=& 
\int d^4 \theta Q^\dagger e^{-V} Q \Phi^\dagger \Phi +
\int d^2 \theta [\Phi^3 (m_0 Q^2 + y_0 Q^3) 
+ \frac{1}{g_0^2} {\cal W}_{\alpha}^2] + {\rm
  h.c.} + {\cal O}(1/M), 
\end{eqnarray}
where $M$ is the higher dimensional Planck scale (which will be somewhat
lower than the four-dimensional Planck scale $M_{Pl}$).
The field
$\langle \Phi \rangle = 1 + \langle F_{\phi} \rangle \theta^2 \sim
\Lambda_H^2/M_{Pl}$, 
 appears here as a supersymmetry-breaking background for the
visible sector. 

We can attempt to eliminate $\Phi$ dependence 
by rescaling  $Q$ according to,
\begin{equation}
\label{rescaling}
Q \Phi \rightarrow  Q \, . 
\end{equation}
Since $Q$ is a physical chiral superfield and $\Phi$ is a background 
 chiral superfield this
transformation respects supersymmetry. The naive result is then, 
\begin{equation}
{\cal L}_{eff}(M) = 
\int d^4 \theta Q^\dagger e^{-V} Q  +
\int d^2 \theta (m_0 \Phi Q^2 + y_0 Q^3 
+ \frac{1}{g_0^2} {\cal W}^2) + {\rm
  h.c.} 
\end{equation}
We see that $\Phi$ only appears beside visible sector mass
parameters, as would
be expected from the conformal coupling
of $\Phi$. The MSSM gauge symmetries  allow only one such mass
parameter, namely  the $\mu$-term.
We will treat the $\mu$ term separately in Section 8 and
assume for now there are no explicit mass parameters in the Lagrangian.

 Classically the $\Phi$-dependence and
 supersymmetry-breaking are again absent in the visible sector after the
 rescaling. However  the
 quantum functional integral measure is not invariant under the
 rescaling, and the resulting $\Phi$ dependence is described by the
 Konishi anomaly \cite{konishi}. 
The most familiar manifestation of this  is the
 well-known axial anomaly, following from the rescaling if we imagined
 the chiral superfield, $\Phi$, to be a pure phase. 

The origin of the
 super-Weyl and Konishi anomalies
 can be understood as follows. We saw above that
 after rescaling,  $\Phi$  appears beside any mass
 terms. Although there are no explicit mass terms, the ultraviolet
 cutoff  $\Lambda_{UV}$  and through it the renormalization scale
 $\mu$ (not to be confused with the $\mu$ term which we are neglecting
 in this section)  always
 provide an implicit mass scale for the theory. In Appendix 2 we
 review how this anomaly arises in supersymmetric 
QED using an explicit supersymmetric
 Pauli-Villars regularization where the regulator
provides an explicit mass term in the Lagrangian. 
More general theories are discussed in
 Ref. \cite{sweyl} and Ref. \cite{gaillard}. 
 Here, we will simply proceed by using the fact that
anomalous  $\Phi$ dependence will multiply  cutoff
dependence.

If we renormalize our cut-off  visible sector theory at an infrared scale
 $\mu$ (that is, we integrate out all modes above some
 experimentally accessible scale $\mu$), 
the Wilsonian effective Lagrangian must take the general
form, 
\begin{equation}
\label{renlag}
{\cal L}_{eff} = \int d^4 \theta Z(\frac{\mu}{\Lambda_{UV} \Phi}, 
\frac{\mu}{\Lambda_{UV} \Phi^\dagger})
Q^\dagger e^{-V} Q  +
\int d^2 \theta y_0 Q^3 
+ \tau(\frac{\mu}{\Lambda_{UV} \Phi}) {\cal W}_{\alpha}^2) + {\rm
  h.c.}.  
\end{equation}
The Lagrangian's dependence on $\mu/\Lambda_{UV}$ follows from
dimensional analysis and the fact that renormalizability of the
theory in which higher dimension
terms have been dropped  means that cutoff dependence can be absorbed in the 
couplings.
 As mentioned above, $\Phi$ 
accompanies  any cutoff dependence. Non-renormalization of the
superpotential coupling $y_0$ implies that cutoff dependence is limited
to the Kahler potential and gauge coupling (which
only renormalizes at one-loop). Since $\Phi$ is formally a
chiral superfield, supersymmetry ensures that $\tau$ is a holomorphic 
function of 
$\Phi$. On the other hand, $Z$ must depend on both the chiral and
anti-chiral superfields, $\Phi$ and  $\Phi^\dagger$.

We can get considerably more information about the anomalous $\Phi$
dependence by relating it to the well-understood chiral 
anomaly in the classical
R-symmetry. Before the rescaling, Eq. (\ref{renolag}) has an exact
formal $R$-symmetry under which $R[\Phi] = 2/3, R[Q] = 0$. (The symmetry
is formal because the lowest component of $\Phi$ is fixed to be one.) This
symmetry is valid even in the presence of mass terms, so it 
 is exact even in the presence of ultraviolet regulator
fields. After the rescaling, we have $R[Q] = 2/3$.
This $R$ symmetry gives us additional information about how $\Phi$
couples in $Z$ and $\tau$ (and in the process we rederive
Eq. (\ref{taushift})).

The fact that the $R$-symmetry must be formally 
exact in the presence of $\Phi$
forces $Z$ to depend only on the $R$-invariant combination, 
\begin{equation}
|\Phi| \equiv (\Phi^\dagger \Phi)^{1/2}.
\end{equation}
Therefore we will consider $Z$ to be a function of $\mu/\Lambda_{UV} |\Phi|$
from now on.

 Now in the absence
of $\Phi$ (that is $\Phi = 1$), the  $R$-symmetry is anomalous.
 An
$R$-symmetry transformation results in a shift in the $\theta$-angle, 
Im$\tau$. When $\Phi$ is fixed to one, $R$-symmetry transformations
will cause $\tau$ to shift anomalously. However, when $\Phi$ is
present, it is rotated by the $R$-transformation and its logarithm will
cancel the anomalous shift to give back an exact symmetry. In
order for the $R$-symmetry to be exact when the $\Phi$ field is
included, the gauge kinetic function $\tau$ evaluated
at the scale $\mu$  must have the form,
\begin{equation} 
\label{tau}
\tau(\frac{\mu}{\Lambda_{UV} \phi})
 = \frac{1}{g_0^2} + 2 b {\rm ln}(\frac{\mu}{\Lambda_{UV} \Phi}).
\end{equation} 
where $b$ is an as-yet determined constant.
 This constant is easily identified
since from the $\mu$-dependence it must be the one-loop supersymmetric 
$\beta$-function coefficient, $b_0$, 
coefficient, 
\begin{equation}
\beta(g) \equiv \frac{d g}{d {\rm ln} \mu} = - b_0 g^3 + ...
\end{equation}
Notice this result has the following interpretation. The $\Phi$-dependent
rescaling of the field has created a cutoff $\Lambda \Phi$. 
To obtain the couplings at a scale $\mu$, we would run according
to the 
renormalization group  from $\Lambda \Phi$ to $\mu$. With
this formula, one can derive the gaugino mass of Eq. (\ref{gauginomass}).

We now proceed to derive the scalar masses. Recall that
 $\Phi = 1 + F_{\Phi} \theta^2$, and
Taylor-expand the ln$\Phi = F_\phi \theta^2$ dependence in
Eq. (\ref{renlag}). First,  
\begin{eqnarray}
{\rm ln} Z(\frac{\mu}{\Lambda_{UV} |\Phi|}) &=& {\rm ln}
Z(\frac{\mu}{\Lambda_{UV}}) - \frac{1}{2} F_{\phi} \theta^2 \frac{d
    {\rm ln} Z}{d {\rm ln} \mu} (\frac{\mu}{\Lambda_{UV}}) + {\rm h.c.} 
\nonumber \\
&+& \frac{1}{4} |F_{\phi}|^2 \theta^2 \overline{\theta}^2 
 \frac{d^2 {\rm ln} Z}{d ({\rm ln} \mu)^2} (\frac{\mu}{\Lambda_{UV}}).
   \nonumber \\
&\equiv& {\rm ln} Z(\frac{\mu}{\Lambda_{UV}}) - \frac{1}{2} \gamma(g, y)
  (F_{\phi} \theta^2 +  {\rm h.c.}) + 
\frac{1}{4} |F_{\phi}|^2 \theta^2 \overline{\theta}^2 (\frac{\partial
  \gamma}{\partial g} \beta_g + \frac{\partial
  \gamma}{\partial y} \beta_y), 
\end{eqnarray}
where we have used the  renormalization group
functions defined by
\begin{eqnarray}
\label{lnZ}
\gamma(g, y) &\equiv& \frac{\partial {\rm ln} Z}{\partial {\rm ln} \mu}
  \nonumber \\
\beta_g(g,y) &\equiv& \frac{\partial g}{\partial {\rm ln} \mu} \nonumber \\
\beta_y(g,y) &\equiv& \frac{\partial y}{\partial {\rm ln} \mu},
\end{eqnarray}
having explicitly isolated the $\Phi$-dependence.
We can take the couplings $g$ and $y$ to be the renormalized couplings
in the exactly supersymmetric theory after $Z$ has been rescaled to
unity. (Recall that because of the superpotential non-renormalization
theorem, the renormalization of the Yukawa couplings is entirely due to
the associated wavefunction renormalizations.)

Let us now compute the soft masses to one-loop order in the visible sector
couplings. We begin by observing that as far as mass terms are concerned 
(see below for $A$-term trilinear scalar couplings) all the terms in ln$Z$ can
 be
rescaled away except for the term proportional to $|F_{\phi}|^2$, by
the superfield redefinition, 
\begin{equation}
\label{Qscaling}
{\rm exp}\{ \frac{1}{2} {\rm ln} Z(\frac{\mu}{\Lambda_{UV}}) -
\frac{1}{2} \gamma(g,y) F_{\phi} \theta^2 \} ~ \Phi \rightarrow \Phi.
\end{equation}
Of course this rescaling also suffers from a formally one-loop anomaly,
but since the transformation parameter is already of one-loop order, the
real anomaly is of higher order and is neglected here. Therefore
having made this transformation, we effectively have 
\begin{equation}
Z(\frac{\mu}{\Lambda_{UV} \phi}) = 1 + 
\frac{1}{4} |F_{\phi}|^2 \theta^2 \overline{\theta}^2 (\frac{\partial
  \gamma}{\partial g} \beta_g + \frac{\partial
  \gamma}{\partial y} \beta_y).
\end{equation}
{}From this the scalar mass-squared renormalized at $\mu$ follows;
\begin{equation}
m_{\tilde{q}}^2(\mu) = - \frac{1}{4} |F_{\Phi}|^2 \theta^2 \overline{\theta}^2 
(\frac{\partial  \gamma}{\partial g} \beta_g + \frac{\partial
  \gamma}{\partial y} \beta_y).
\end{equation}
An important feature of this result is that the scalar mass squared
only arises at two-loops. As discussed in Section 2, this
is a very desirable property of the spectrum.
Now to
 compute scalar masses to one-loop order we must clearly
retain  terms up to two-loop order
in  the scalar mass-squared. Therefore each of $\gamma$, $\beta_g$ and
$\beta_y$ must be kept to one-loop order. They have the general forms, 
\begin{eqnarray}
\label{rg1loop}
\gamma &=& c_0 g^2 + d_0 y^2 \nonumber \\
\beta_g &=& - b_0 g^3 \nonumber \\
\beta_y &=& y (e_0 y^2 + f_0 g^2).
\end{eqnarray}
{}From this it follows that the scalars have mass-squareds
\begin{equation}
\label{scalarmass}
m_{\tilde{q}}^2 =  \frac{1}{2} \{ c_0 b_0 g^4 - d_0 y^2 (e_0 y^2 + f_0 g^2) \} 
|F_{\Phi}|^2.
\end{equation}

{}From eq. (\ref{scalarmass}) we can see a general feature of
anomaly-mediated supersymmetry breaking. Since for all gauge
representations, $c_0 > 0$, the contribution to scalar mass-squareds from
purely gauge field loops is positive for asymptotically free gauge
theories and negative for infrared free gauge theories. This poses a
phenomenological danger for scalars whose visible sector couplings are
dominated by infrared free gauge interactions. Other sources of
supersymmetry-breaking must be
sufficiently large to counter
the negative contributions found here.
We will discuss these contribtuions 
in the following section.

Second, we observe that the superfield transformation, Eq. (\ref{Qscaling}), 
induces trilinear $A$-terms which, using the superpotential 
non-renormalization theorem, are
  proportional to the $\beta$ functions of the corresponding Yukawa 
couplings.\footnote{This is corrected from our
original version and confirms the result of  Ref. \cite{markus}.} In order 
to avoid confusion we will restore indices $i, j, k, ...$ labelling the 
various chiral 
superfields. Then we find,
\begin{equation}
A_{ijk} = \frac{1}{2} (\gamma_i + \gamma_j + \gamma_k) y_{ijk} F_{\Phi},
\end{equation}
where each $\gamma_i$ has the form given in Eq. (\ref{rg1loop}).

We wish to discuss a few general features of the derivation
of the gaugino and scalar masses. First of all, although the
results we derived here all occur at loop level, they should be
thought of as the initial conditions for a renormalization group
running in the MSSM. We have not yet done renormalization group
running {\it below} any SUSY threshold. The final low-energy
spectrum would include these effects as well. However, this running
is not from the Planck or compactification scale, but only
from the scale of a SUSY breaking mass. Other 
running has already been included.

Second, our results appear similar in spirit to those of Refs. 
\cite{wavefunction} \cite{analytic}, 
who computed higher loop masses in terms of one-loop 
parameters
of the beta function and anomalous dimension in the context of gauge-mediation.
This similarity can be understood as follows. In gauge-mediation,
the messengers of supersymmetry breaking are vector-like multiplets which
appear in loops. In our calculations, the ``messengers'' of supersymmetry
breaking can be thought of as the regulator fields. 
This interpretation in terms of loops of regulator fields
is  made explicit in Appendix B.
This leads to quite distinct results however. For example,
the beta function dependence in the gaugino and  scalar mass squared
arises because not only charged matter, but gauge
bosons themselves have regulator fields which
directly couple to supersymmetry breaking. This
leads to a very different spectrum from gauge-mediated models.
 However,
 from the 
vantage points both of diagrammatic perturbation theory and the
more formal analysis presented above, it can
be useful to use the  similarity of the 
results.
For example, we can understand the sign of the
soft scalar mass squared. 
Because the regulator fields yield loops of the opposite
sign, the sign should be the opposite of that given
by the physical messenger fields, which is to say the opposite
sign of the beta function.
 This analogy is  also useful in understanding
how loops are cut-off at a high scale, where the supersymmetry
violating regulator mass goes away.

Finally, let us consider the effect of having supersymmetric mass
thresholds, such as a GUT scale, in the visible sector. 
As discussed earlier, the dominant 
tree-level effect is that the massive fields will aquire supersymmetric
splittings because their mass parameters in the lagrangian, $m_0$, 
will be multiplied by $\Phi$ after the rescaling. However, assuming
that $m_0$ is much larger than any experimentally accessible energy, we
are really interested in how the supersymmetry-breaking  splittings
among the massive states feeds down radiatively 
to the light visible sector fields. The central
point is that once we integrate out the massive fields, the effective
Lagrangian for the light fields must be of the general form of
eq. (\ref{renlag}), with the substitutions, 
\begin{eqnarray}
Z(\frac{\mu}{\Lambda_{UV} |\phi|})
&\rightarrow& 
Z(\frac{\mu}{\Lambda_{UV} |\phi|}, \frac{m_0}{\Lambda_{UV}}) \nonumber \\
\tau(\frac{\mu}{\Lambda_{UV} \phi}) &\rightarrow&
\tau(\frac{\mu}{\Lambda_{UV} \phi}, \frac{m_0}{\Lambda_{UV}}). 
\end{eqnarray}
We see that the new $m_0$ dependence does not enter at all into the
considerations that determined the soft masses in terms of the {\it
  renormalized} couplings at $\mu$, since these only involved
derivatives with respect to $\mu$. The only effect of the $m_0$
threshold is that it
changes the renormalized couplings themselves as functions of the
ultraviolet couplings above $m_0$. Since we directly measure and work
in terms of the
infrared couplings and are ignorant of the ultraviolet ones, the
presence of the $m_0$ threshold is irrelevant for our
soft mass calculation. This provides a sort of ultraviolet ``immunity''
for our results. There are corrections to this picture coming from
possible non-renomralizable terms in the effective lagrangian,
suppressed by $m_0$. If $m_0$ is very large then these effects can be
neglected, but clearly this would not be the case if there were a
supersymmetric threshold close to the weak scale. Diagrammatically,
the irrelevance of heavy states can be understood in terms of a cancellation
between loops of massive states and their regulators,
which both have the same coupling to the supersymmetry breaking $\Phi$ field.

\section{Additional Contributions to the Soft Scalar Masses}

In the previous section, we evaluated the anomaly-mediated
contribution to the gaugino and scalar masses. We found
the remarkable fact that although gaugino masses arise
at one-loop, the scalar mass squared only arises at two.
However, we also showed  
that the spectrum had a serious problem, namely the negative
mass squared for the sleptons.  However, there are other ways
scalars can obtain masses; once supersymmetry is broken,
one expects this to be communicated at the radiative
level  to the scalar mass squared
since no symmetry prevents such a mass. In this section,
we will see that in the presence of additional bulk fields,
 there are indeed  additional contributions to the
slepton (and squark) masses.  
 In Section 10, we will describe
the requirements for these to be of the type and order of magnitude
to generate a reasonably natural spectrum consistent with flavor-violating
constraints. 

\subsection{Bulk Radiative Corrections}

Let us reconsider our higher dimensional scenario with some number and
types of bulk fields with non-renormalizable Planck-suppressed 
 couplings to the visible and hidden sector fields. 
Examples of these are
 bulk scalars coupling (in higher dimension operators) to
both visible and sequestered sector fields,  3-brane fields
 appearing in the gauge kinetic functions of bulk gauge fields, or
supergravity fields themselves. 
We assume that there are
no couplings of the bulk to the visible sector which give rise to
renormalizable strength interactions (after compactification). So for
example we exclude consideration of 3-brane fields which are
charged under the bulk gauge fields, or Yukawa couplings of bulk and
3-brane fields. If the fundamental theory contained any such
couplings it is assumed that the associated bulk fields have
Planck-scale masses and have decoupled. 

Let us begin by considering the form of radiative corrections to the
soft masses from bulk loops, within the framework of the 
four-dimensional effective theory below $\mu_c$.  
Here it is very important that we  consider only
the regime  $\mu_c<M_{Pl}$.
Unlike conventional supergravity, where everything is confined
to a single four-dimensional spacetime, 
there will be a well-defined perturbative loop
expansion when computing supersymmetry-violating effects with 
a well-defined cut-off for divergent integrals.  This
follows   from the assumption of couplings suppressed by the Planck
scale and dimensional analysis. Here we show explicitly how this
arises.

It might appear curious that using higher dimensions has actually
{\it improved} the convergence of supergravity. This is of
course  true only for supersymmetry-violating operators that 
require communication across the compact dimension.  
To actually compute these corrections, one can use the results
of a conventional supergravity calculation using our special Kahler
potential. However, it is edifying to understand the loop calculation
directly in position space in the higher dimensions and before Weyl rescaling. 
We consider here the case with one extra dimension for concreteness.
 One can then see directly that the  separation
$r_c \sim 1/\mu_c$  effectively acts as a
(physical) point-splitting regulator for the one-loop corrections and
therefore one obtains a finite result. This is most easily seen by
considering the form of a  one-loop corrections in position-space.
We assume that the bulk fields  contain only  Planck mass suppressed
couplings to both the visible and sequestered sectors.  In the
case of the gravity multiplet, a loop diagram would require
two Planck mass suppressed vertices with two derivatives.
A bulk scalar would couple to matter in the form
\begin{equation} \label{higherord}
{\cal S} \supset \int d^5 x \int d^4 \theta \left( c_1{ \Sigma^\dagger 
\Sigma B B^\dagger
\over M_{Pl}^2} \delta(x_5-r_c) + c_2{Q Q^\dagger B B^\dagger \over M_{Pl}^2} 
\delta(x_5) \right)
\end{equation}
where $B$ is a bulk field.  $M$ (as opposed to
$M_{Pl}$) denotes the five-dimensional Planck scale.
Note that we have taken quite a general form for the coupling
between the bulk and sequestered and visible sectors, assuming
only that it was suppressed by $M_{Pl}$. One could
in principle incorporate also single $B$ couplings if $B$
is a singlet; however these will only contribute to irrelevant
derivative couplings.

 The above interaction gives rise  to  a two-derivative
coupling of $B$ to $Q$. We therefore  expect a loop-integral
 of the form
\begin{eqnarray}
m_{\tilde{q}}^2 &\sim& \frac{1}{M^3} \int d^4 x
\partial_{\mu} G(x, r_c)  {\rm Str} \Delta(x, r_c) \partial_{\mu} G(x, r_c),
\end{eqnarray}
where, $G$ is the five-dimensional massless 
scalar Euclidean Green function which
connects the bulk interaction with $\tilde{q}$ at the spacetime origin
to any point on the hidden sector 3-brane at which there is a
supersymmetry-breaking interaction, $\Delta$,  which can split 
the  masses of the bulk modes in the four-dimensional effective theory.
The derivatives arise from the higher dimension
operator which couples the bulk fields to the visible
sector and $\Delta$ could be a supersymmetry
breaking mass-squared insertion for the bulk field for example.
  Since the bulk
couplings to the hidden sector are Planck-suppressed (by powers of $M$ in 
five-dimensions) we have,
\begin{equation}
\langle \Delta \rangle \sim {\cal O}(\frac{F_{\Sigma}^2}{M^3}).
\end{equation}
Since $m^2_{bulk}\sim F_\Sigma^2/M_{Pl}^2$ and $M_{Pl}^2 \sim r_c M^3$
we find,
\begin{equation}
{\rm Str} \langle \Delta \rangle \sim  r_c {\rm Str} m_{bulk}^2.
\end{equation}

Now an
ultraviolet divergence would normally occur when the two interaction
points coincide. However in the present case
 they cannot get closer than $r_c$ and therefore
the $x$-integral is finite, and by dimensional analysis 
is of order $\mu_c^4$. Our final result is then,
\begin{equation}
m_{\tilde{q}}^2 \sim \frac{{\rm Str} m_{bulk^2}}{16 \pi ^2 M_{Pl}^2}
\mu_c^2 \sim \frac{|F_{\Sigma}|^2}{16 \pi ^2 M_{Pl}^4} \mu_c^2. 
\end{equation}
This is the key point which we emphasized in the introduction;
loops are finite and cutoff by $\mu_c$.

One can  generate small $A$-terms from the bulk as well.

If we now assume $F_\phi \sim F_\Sigma/M_{Pl}$, we
see that a positive slepton mass squared requires $\mu_c/M_{Pl} \sim 1/10$. 

One can obtain the exact result of this calculation
by substituting our special form of the Kahler
potential including the additional contributions
from Eq. (\ref{higherord}) using a physical
cutoff \cite{gaillard} of the compactification scale. 

In a Scherk-Schwarz scenario, the authors of
Ref. \cite{gc2} computed precisely the above
loop effects and found 
\begin{equation}
m^2_{\tilde{q}} \sim  0.88 
 K^{-1 ~ Q Q^\dagger} \left({\cal R}_{Q^\dagger Q} -
K_{Q^\dagger Q} \right) {\mu_c^4 \over M_{Pl}^2}
\end{equation}
where we have taken $m_{3/2}=\pi/2 \mu_c$.
Notice in this scenario, the bulk mass is not Planck-suppressed.
So the analogy to the above formula requires taking $m_{bulk}\sim F_\phi 
\sim \mu_c$.
We then find the compactification scale should be given
 as $\mu_c/M_{Pl} \sim 1/100$.

There are a few further observations  about this result. The
second term, representing the pure gravity contribution,
is negative. This means there {\it must} be additional
bulk contributions.

A particularly interesting such case is if the bulk scalars
arise from the dimensional reduction of higher dimensional
supergravity.  This is interesting because such bulk contributions,
like those of gravity, can be flavor-blind. For example, in the Horava-Witten
setup, the modulus which determines the scale of the eleventh
dimension has flavor-blind couplings. One can compute the contribution
from a single such modulus; the result is that any individual such
scalar gives a contribution 1/3 the size of gravity. Therefore,
with at least four such scalars, one can obtain a positive
flavor-independent bulk contribution to the soft scalar mass squared.

Another interesting thing to observe from the above formula is
the preferred choice of compactification scale, which is approximately
$0.01 M_{Pl}$ in order to compete with the loop-induced anomaly-mediated
negative slepton mass squared.  Depending on the number
of dimensions of size $r_c$, one finds the higher dimensional
Planck scale ranges from about $3\cdot 10^{16}$ GeV to $2\cdot 10^{17}$ GeV,
where we have varied the number of large dimensions from one to six.
So we see this allows for the possibility of unifying 
  gauge and gravitational couplings,
while giving a consistent positive flavor-degenerate contribution
to the soft scalar masses! Of course in this case there is only
a small separation between the compactification and string scales.
Nonetheless these simple dimensional analysis estimates
indicate a very intriguing possibility.

We observe that this is different from the strongly coupled heterotic
string setup for a couple of reasons. One is that there
were two distinct scales in that case: the Calabi-Yau radius
and the 11-dimensional radius. That can perhaps
be accomodated; we assumed common radii for simplicity.
Furthermore, one can only analyze the heterotic string
in the weakly or strongly coupled limits; in the
regime of intermediate string coupling, the scales
could be different. Furthermore, there can be
additional bulk states which can contribute
to the soft scalar masses.
The more problematic aspect is that flavor is established
geometrically; this means that some  scalars arising
from the Kaluza-Klein reduction of the gravitational multiplet
would not be exected to
have flavor-blind couplings. The additional difference 
in the specific phenomenology which has been done is
the assumption that the dilaton is present in the low-energy theory;
this gave very different contributions to the scalar and gaugino
and cannot address the flavor problem.
With the dilaton present, there are  direct {\it tree-level}
couplings in the
four-dimensional effective theory
between the supersymmetry breaking and visible sectors.

We close this section with an aside. One might
have thought one can readily generate a flavor-blind
contribution to the soft scalar mass squared
by the one-loop correction to the vacuum energy, which
would require a different constant superpotential
term than that determined at tree-level. In fact,
such loop corrections have been considered
in Ref. \cite{kim, brignole}.  
However, we see no such corrections before Weyl rescaling. The only
bulk radiative corrections to the soft scalar masses involve
the direct couplings of the bulk fields to the visible sector fields.
This follows from the general analysis of Section 4. There  we saw
the apparent contributions to the mass squared from $W$ must and
did cancel (which was more apparent before integrating out
the auxiliary field and Weyl rescaling). This is still true
when one-loop corrections are included. The contribution from Weyl rescaling
which appears in the kinetic term
cancels against the potential contribution. 
Now one has the option of rescaling the Planck mass to eliminate
the higher dimension kinetic terms; this then puts the
additional contribution in the Kahler potential as higher
dimension terms. Therefore, additional contributions
associated with vacuum energy which are enhanced
by the number of chiral multiplets really arise from an assumed
flavor-independent higher dimension operator in the Kahler
potential, of the form $|Q_i^2|^2/M_{Pl}^2$. It is worth
noting that the large one-loop mass correction of 
\cite{kim, brignole}
relies on this assumed form of the Kahler potential. In 
our case, loops of matter
fields give too small a contribution to the vacuum energy in our
case since the mass arose only at one-loop.
This means
that only bulk fields that couple directly to matter in higher
dimension terms
give a contribution to the scalar mass squared.

\section{ The $\mu$ Problem}

No theory of supersymmetry breaking would be complete without a solution
to the $\mu$ problem. Although the $\mu$ term in the superpotential
preserves supersymmetry, we know the scale is
essentially the same as that of supersymmetry
breaking so is presumably connected to it. Therefore one expects
the mechanism which breaks supersymmetry to also be responsible
for inducing the $\mu$ parameter. 

The lack of solution to the $\mu$ problem is one of the least
satisfying aspects of gauge-mediated supersymmetry breaking.
Although solutions exist \cite{mu1} \cite{mu2} \cite{mu3} \cite{mu4}, 
they are not compelling.
The basic problem is that in gauge-mediated models,
supersymmetry breaking is induced at loop-level. So a term
in the effective superpotential which induces $\mu$ is generally
too large without some additional suppression factor;
in this case $\mu B$ is too large. For example,
if there is a singlet $S$ which has nonvanishing auxiliary component,
one can write $\epsilon \int d^2 \theta S H_u H_d$. If $\epsilon \sim 10^{-2}$,
one can successfully obtain a $\mu$ parameter of the correct size.
But then $\mu B$, the scalar mass squared term, is too large. 
On the other hand, there is a simple solution in  hidden sector models;
one can construct a term in the Kahler potential \cite{gm}
$\Sigma H_1 H_2/M_{Pl}$.

In our models, we can clearly not simply
fine-tune the problem away. If we permit a Peccei-Quinn (PQ) symmetry-breaking
term $\int d^2 \theta H_1 H_2 \Phi^3$ (where we have included
the essential dependence on $\Phi$), one finds that $\mu B$ is
far too large. However, this problem might be readily solved
by a miraculous cancellation if it is assumed the $\mu$
term arises from the following higher dimensional term in the 
Kahler potential:
\begin{equation} \label{muterm}
{\cal L}_{mu-term} 
 = \alpha \int d^4 \theta {1 \over M_{Pl}} \left( \Sigma + \Sigma^\dagger
\right)  H_1 H_2 \Phi^\dagger \Phi
 +{\rm  h. c. }
\end{equation}
 Here $\Sigma$
is the supersymmetry breaking hidden sector field. After the rescaling,
Eq. (\ref{rescaling}), (applied only to the visible sector fields) one
obtains, 
\begin{equation} \label{muterm2}
{\cal L}_{mu-term} 
 = \alpha \int d^4 \theta {1 \over M_{Pl}} \left( \Sigma + \Sigma^\dagger
\right)  H_1 H_2 \frac{\Phi^\dagger}{\Phi}
 +{\rm  h. c. }
\end{equation}
Notice that with this operator, there is no {\it classical} contribution to 
 $\mu B$ when  $F_\phi \propto F_\Sigma$. It is straightforward to see that 
there is an anomaly-mediated 
contribution however resulting from Eq. (\ref{Qscaling}) given by
\begin{equation}
B = \frac{1}{2} (\gamma_{H_1} ~+~ \gamma_{H_2})~ F_{\Phi}.
\end{equation}

One can ask where such an operator as Eq. (\ref{muterm}) arises, and if there
are couplings of $\Sigma$ which produce other operators
that do generate a $B$ term. We provide a simple example, with a
five-dimensional fundamental spacetime for concreteness.

Assume there is a massive bulk vector field $V$ with the following four
dimensional Lagrangian
\begin{equation}
{\cal L}_V=\int d^4 \theta r_c m^2 V^2 + aV (\Sigma + \Sigma^\dagger) M^{1/2}
+{b \over M^{1/2}} V H_1 H_2 + \ {\rm h.c.}
\end{equation}
Here, $M$ is the Planck scale of the five-dimensional theory
(this can however be generalized to higher dimensions), $a$ and $b$
are numbers, and $m$ is the vector mass in the five-dimensional theory.
Notice that we have written the most general  term allowed
at this order in a $1/M$ expansion.

One can then work out the contribution to the operator above 
Eq.(\ref{muterm}). 
It scales like $a b/r_c m^2$. For $m$ of order the compactification
scale and $a$ and $b$ somewhat less than one, one finds an
acceptable $\mu$ parameter.

This model serves as an existence proof. It is probably
not as complicated as the generation of a $\mu$ term
in gauge-mediated models, but is not as straightforward
as the mechanism in hidden sector models with no additional
constraints. One could construct other models; the ones
we have so far constructed are less natural in that
they would permit additional couplings and one would
need to assume relations among parameters. 
Since we have not made very unreasonable
assumptions, we expect that a fundamental high energy theory
might allow for a vector field of the type we have described.
However, there might be a yet more compelling solution
to the $\mu$ problem which we have overlooked.

\section{CP Violation}
We claim this scenario is also very advantageous from
the vantage point of CP violation.
There are two types of potential CP problems \cite{ratnir}. The first is
the SUSY strong CP problem which requires the phase of the $A$
and $B$ parameters to be small, of order $10^{-2}$ \cite{ratnir}. Clearly,
this problem is solved if $A$ and $B$ are only radiatively generated.
The existing phases can be rotated away. In our models, the
problem with the phase of $A$ is automatically solved. The
problem with the phase of $B$ depends on the precise
solution to the $\mu$ problem. In our
example, the problem is solved.

The other CP problem is that $\epsilon_K$ can be too large \cite{ratnir}. This
problem is distinct and tied up with the solution to the flavor problem.
In the case that squark masses are degenerate, there is clearly
no problem. 
Furthermore, in a higher dimensional theory, it is possible
that there are no operators which communicate CP violation between our 3-brane
and the bulk. Then it could be natural for the SUSY breaking sector to preserve
CP.

\section{Mass Spectrum and Phenomenology}
 
In this section we will assemble and examine our results for the soft
breaking terms in the MSSM. These will be determined by the anomaly-mediated
contribution and an unknown bulk contribution.
 We will work in the approximation that all 
Yukawa couplings vanish except for $y_t$ and $y_b$. 

We first derive the anomaly-mediated contribution to the 
masses, which will dominate for all but the sleptons and Higgses.
We remind the reader that although the formulae below are
loop-level results, they should be considered as {\it matching}
conditions for a full renormalization group analysis in the
MSSM below the superpartner mass scale.

Recall, the formula for the anomaly-mediation
contribution to the  soft scalar mass squared
\begin{equation}
m_{\tilde{q}}^2=-{1 \over 4} |F_\phi|2 \left( {d \gamma \over dg} \beta_g
+{d \gamma \over dy} \beta_y \right)
\end{equation}
where $\gamma \equiv d log Z(\mu)/d log \mu$.
When we take
$\gamma=c g^2+d y^2$, $\beta_g = -b g^3$,
we can work out the masses of the squarks and gauginos
in terms of the above parameters.

We then have for the anomaly-mediated contribution
for $\tilde{q} =   \tilde{t}_R$ and the $H_u$ doublet of scalars, 
\begin{equation}
m_{\tilde{q}}^2 =  \frac{1}{2} \{ c b g^4 +{d\gamma \over dy_t} \beta_t \} 
|F_{\phi}|^2.
\end{equation}
Similarly, for $\tilde{q} = \tilde{b}_R$ and the $H_d$ doublet of scalars,
\begin{equation}
m_{\tilde{q}}^2 =  \frac{1}{2} \{ c b g^4 +{d \gamma \over dy_b} \beta_b \} 
|F_{\phi}|^2,
\end{equation}
and for $\tilde{q}=\tilde{t}_L, \, \tilde{b}_L$,
\begin{equation}
m_{\tilde{q}}^2 =  \frac{1}{2} \{ c b g^4 +{d \gamma \over dy_t} \beta_t
+{d\gamma \over dy_b} \beta_b \} 
|F_{\phi}|^2.
\end{equation}
For the remaining scalars, 
\begin{equation}
m_{\tilde{q}}^2 =  \frac{1}{2} c b g^4  |F_{\phi}|^2.
\end{equation}
It should be understood in the equation above that the gauge
terms indicate the sum over the appropriate terms according
to the gauge charge of the scalar; the Higgs and top squark
mass are of course not the same.

The numerical value of $y_t$ and $y_b$ is uncertain because of
the unknown value of tan$\beta$.
For simplicity, we
 give the values of the masses for vanishing $\beta_t$ and $\beta_b$.
We take the values of the parameters at about 1 TeV from Ref. \cite{jones}:
$\alpha_Y=0.01$, $\alpha_2=0.032$, $\alpha_3=0.1$.
With the well-known renormalization
group coefficients \cite{jones},  we find the scalar masses
\begin{eqnarray}
m_{sleptons}^2 &=& - 1.3 \times 10 ^{-5} |F_{\phi}|^2 + m_{bulk_l}^2 
\nonumber
\\
m^2_{squarks}  
&=& 5.5 \times 10 ^{-4}  |F_{\phi}|^2 + m_{{bulk}_H}^2
\nonumber \\
m_{H}^2 &=& - 1.3 \times 10^{-5}  |F_{\phi}|^2 + m_{{bulk}_H}^2
 \\
\end{eqnarray}
while the gauginos acquire soft mass-squareds are 
\begin{eqnarray}
m^2_{gluino} &=& 6.1 \times 10^{-4} F^2_{\phi} \nonumber \\
m^2_{wino} &=&  6.4 \times 10^{-6} F^2_{\phi} \nonumber \\
m^2_{bino} &=&  7.0 \times 10^{-5} F^2_{\phi}. 
\end{eqnarray}

Even without knowing the bulk contributions of $F_\phi$, there are several
interesting features of the spectrum which we can identify.

\begin{itemize}
\item The ratio of gaugino masses is determined. We find
$m_3:m_2:m_1=3.0:0.3:1$.
\item The wino/zino are the lightest of the gauginos. 
Furthermore, assuming a bulk contribution to the slepton mass
at least of order of magnitude of the bulk contribution,
the wino/zino are the lightest supersymmetric particles!
\item The squarks and gauginos are the heaviest particles,
and are quite heavy; they are an order of magnitude
heavier than the wino.
\end{itemize}
In fact, the wino and zino are nearly, but not exactly,  degenerate. Refs. 
\cite{gunion} 
have studied some of the phenomenology of such a situation,  based on the 
string-derived models of Ref. \cite{lust}. 
In the approximation $M_2 < M_1 < \mu$, one can approximately
determine the mass difference \cite{martin} \cite{pol} \cite{kane}
 to find
\begin{equation}
m_{\tilde{w}}-m_{\tilde{z}}\sim{m_2\over 2} \left( {m_w \over \mu} \right)^4
\end{equation}
This is fascinating. The wino and zino are so nearly
degenerate that the wino lifetime should be quite large.
Of course this depends on parameters, but for reasonable
parameters, one finds the lifetime of the
wino is comparable to that of a muon, since the lifetime
scales as $(\Delta m)^5$. This means supersymmetric events
should have a very striking signature. Sometimes the LSP will
manifest itself as a zino, and look like a ``typical'' LSP. However,
sometimes it will appear as a wino. The wino event will not deposit
energy in the calorimeter, appearing
as a missing energy event,
 but the wino will be detected in the muon chamber.
With good time of flight, one should even be able to learn about the mass.
In any case,  this striking signature of these events should be sufficient to
identify the sequester-sector scenario.

There are a couple of comments on the spectrum so far. First,
away from the fixed point, 
$H_u$ can have a  largish contribution which is a function of $y_t$,
and $H_d$ could also have a largish contribution if $y_b$ is
large.  However, these contributions are proportional to $\beta_{y_t}$
and $\beta_{y_b}$. There is a large $\tan \beta$ fixed point \cite{irfp}.
If the world sits at the fixed point of $y_t$, the extra
contribution to the $H_u$ mass is reduced to zero. Similarly, the
$H_d$ mass contribution is very sensitive to $y_b$ near the fixed
point. So although the Higgs mass seems to need some tuning,
this is not the case if one is at a fixed point. However, since
we have less parameter freedom in our model than
would be allowed in a general hidden sector, one would have to check
the consistency of the whole picture. This is an interesting
problem for future study.

The second comment  is the fact that the stop squark mass
is large. Naturalness bounds constrain the stop mass since
it feeds directly into the Higgs mass; large stop implies
large Higgs mass. However, these masses are being
given at a low scale, the supersymmetry breaking scale of order 1 TeV.
Below this scale, one would apply the standard renormalization to
determine the spectrum. Since we are starting the scaling
at such a low energy scale, there is no large logarithm. This means
the naturalness constraints on the stop mass (and the gluino mass)
are less severe. Masses of 2 TeV are perfectly consistent. In
fact, one might even be willing to accomodate heavier masses;
if true, this could be discouraging from the point of view of
finding these superpartners. The upshot is that the naturalness
bounds and reach of the collider needs to be reconsidered
in light of our different spectrum.

Finally, we comment on the gaugino mass spectrum. It is commonly
understood that a necessary consequence of gauge unification
is that the gaugino mass parameters  associated with SU(3), SU(2), and U(1)
must arise in the ratio $g_3^2:g_2^2:g_1^2$.  However, our scenario
is perfectly consistent with unification. One way to think of why
this happens is that there are large threshold corrections. The
heavy GUT mass particles receive tree-level supersymmetry breaking,
whereas the light states only receive the anomaly-mediated one-loop
contribution. So the corrections from integrating out the
heavy GUT states are so large as to give the prediction
we have made here.

Notice that the gaugino masses depend on the beta functions. So the
light wino is understood in terms of the fact that SU(2) scales relatively 
slowly.

We now  consider the size of bulk effects. These 
are more model-dependent, so we simply
constrain them by the necessary
phenomenological criteria.  The first
essential requirement is that bulk contributions  are sufficiently large
to make the slepton mass squared positive.  This means
that they should be at least of order $10^{-5}|F_\phi|^2$.

The second requirement is that flavor violation is sufficiently
suppressed.  There are two possibilities; one is that sleptons
are sufficiently heavy to suppress flavor violation, as proposed in
Ref. \cite{andy}.
This would require slepton masses of order  $1-10$ TeV.  The other possibility
is that the sleptons are relatively light but sufficiently
degenerate to suppress flavor violation, due to flavor-conserving
bulk couplings.

Now one might expect a comparable bulk contribution to the squark
mass squared. However, with no flavor symmetry requirement on the
bulk contribution, an additional bulk contribution of order 1 TeV
would be far too large, and introduce unacceptable flavor violation
into the squark mass matrix. So it is clear that the bulk
states that couple to quarks must give a significantly smaller
contribution than their leptonic counterparts. This is naturally
obtained with the ``switch'' described in Section 5. If the
states coupling to squarks have mass in excess of the compactification
scale, their contribution to the squark mass will exponentially decouple.
One other potentially unsatisfactory feature of the heavy slepton
scenario is that one might also expect the Higgs to then have
a relatively heavy mass, decreasing the naturalness of the scenario.

The second possibility is that the sleptons are light. Then
they must necessarily be fairly degenerate. 
Recall that this could happen if the light bulk
moduli arise from the Kaluza-Klein reduction of the 
gravitational multiplet as discussed in Section 7.
One
would then expect a small correction to the anomaly-mediated
contribution to the squark mass matrix (recall the large QCD-induced
contribution to the squark masses). 

Notice that in both scenarios, under the assumptions outlined
above, the squark mass is dominated by the anomaly-mediated contribution.
So we have one more prediction assuming small slepton mass which is 
\begin{itemize}
\item The ratio of gaugino to squark mass is $1.05$.
\end{itemize}

To give a general idea of what these spectra look like, we
present two ``typical'' examples. In the first, the light
chargino is near the experimental bound, and in the
second, the squarks and gluinos are towards the upper limit
of the experimentally detectable range at the LHC.

Spectrum I: $m_{gluino}=980$ GeV, $m_{wino}=89.974$ GeV, $m_{``zino''}=
89.348$, $m_{``bino''}=330$ GeV, $m_{squark}=930$ GeV, $m_{slepton}=200$ GeV.
Here we took $\mu=250$ GeV. For this value,  $\Delta m \equiv
m_{wino}-m_{``zino''}=600$ MeV. If, on the other hand,
one takes $\mu=800$ GeV, one finds $\Delta m=9 $ MeV..

Spectrum II: $m_{gluino}=1950$ GeV, $m_{wino}=194.23$ GeV, $m_{``zino''}=
194.112$ GeV, $m_{``bino''}=660$ GeV, $m_{squark}=1850$ GeV, 
$m_{slepton}=300$ GeV. Here we took $\mu=500$ GeV, which gives
$\Delta m=120$ MeV. For $\mu=1600$ GeV, one finds $\Delta m <10$ MeV.

The values for $\mu$ were motivated by a positive higgs mass squared
with and without the Yukawa-dependent contribution. We have
not included these to distinguish the top and bottom squark masses,
but away from the fixed point, one should.

As emphasized above, a remarkable feature of the spectrum
is the light wino and the small mass splitting from the zino.

In fact, these splittings have not included further radiative corrections.
Particularly for very small splitting, custodial SU(2) violations
should modify these numbers. We expect that this corrects
our result at the level of $0.1\%$ of the wino mass, but a calculation
is necessary to confirm this.

The lifetime of the wino, for small mass splitting from the zino,
is
\begin{equation}
\tau =\left({ m_\mu \over \Delta m} \right)^5 8 \times 10^{-8} \ {\rm sec}
\end{equation}

If we take $\beta \gamma$ of order unity, and assume a lifetime
of 10 nsec is required to get to the muon chambers, this translates
into a mass splitting of about 160 MeV in order
to obtain our ``smoking gun'' signature. It is clear that
the mass splitting is in a very interesting range. Much of the time
$\Delta m$  is small compared to this number, or within a factor of 10, in
which case one might hope to see a track sufficiently distinctive
from a tau to distinguish it. Only for small $ mu$ and $m_{wino}$
was the liftime sufficiently short to decay before reaching
the muon chambers. Clearly, a more detailed analysis is in order
to assess the reliablity of this signature over the allowed
parameter range.

It might seem remarkable that our spectrum is so predictive.
The reason is our assumption that there are no direct couplings
between the supersymmetry breaking and visible sectors. It
is true that this assumption is not necessarily associated
with the existence of higher dimensions. In fact, at one-loop,
the anomaly-mediated contributions to both scalar and gaugino
masses will arise in {\it any} hidden sector model. However,
without the assumption of higher dimensions, the nonexistence
of tree-level masses, at least for the scalars, would be entirely
unnatural. What we have is a mechanism for naturalness which is
not implicit in the low-energy theory. We therefore expect
this spectrum is associated with the existence of higher dimensions.
Without this assumption, one would find an unacceptable hierarchy
between the gaugino and scalar mass.   This gives the amazing
possibility of testing directly for the existence of higher dimensions
in the mass spectrum of supersymmetric partners.
If we are not unlucky, the wino will be ``long-lived'' and
there will be the ``smoking gun'' signature described above.
This would be later confirmed by a measurement of the spectrum.
If the wino decays before the muon chamber, it will look tau-like
and might therefore be difficult to identify. However, measurements
of the spectrum will always  serve to confirm or reject this
scenario.

\section{Conclusions}

Clearly physics permits new possibilites if we live
in a higher dimensional universe. One important
feature from the low-energy point of view is that
the notion of naturalness gets extended. Not only
symmetries, but also the geometry of space-time can forbid
operators from appearing in the low-energy Lagrangian.
Many of the standard problems with hidden sector
models have new potential resolutions in this framework.

We point out that our model is perfectly consistent with
grand unification of gauge couplings.
However, whether or not we can unify the gauge and gravitational
coupling is more model-dependent. Since we have a preferred
value of the compactification scale to generate scalar
masses, one is not necessarily free to choose the mass
scales to provide this further unification. However,
we have seen in Section 7 that it could be possible
for the scales to permit gauge-gravity unification. Given
the uncertainty in the estimates, we find this an intriguing
possibility.

Amazingly, the asssumption of a sequestered sector is extremely
predictive. The spectrum is different from any other model
which accomodates gauge unification. This is due to the
large threshold 
corrections when crossing a heavy particle mass scale.
We have found predictions for the ratio of gaugino masses,
and the ratio to the squark mass. We also have pointed
out that the small mass splitting of the two lightest
superpartners permits the potential for a dramatic
signature. One might question the direct connection
between these signatures and the assumption of higher dimensions;
most follow from the gaugino mass contribution present in {\it any}
hidden sector. However, it is difficult to envision any other
hidden sector which naturally gives scalar masses of the same
order of magnitude without arbitrary and unnatural  assumptions.
Furthermore the ratio of gaugino to squark mass should be unique to
our scenario; any other scenario should give additional contributions
to the squark mass.

Our assumptions are quite general, but there are some requirements
for the high energy model. There must be at least one
higher dimension which is stabilized at a mass scale near but
below the Planck scale. Bulk fields which
are lighter than the compactification scale  are required to provide
the positive contribution  to the slepton mass squared. Furthermore,
the light states should not introduce large flavor dependence
in the squark or slepton mass squared. Clearly it is
difficult to assess the likelihood of these assumptions
without a better understanding of the mechanism of moduli stabilization.

There is much which remains to be understood. In a future
paper, we will present the effective theory of particular
sequestered sectors, in particular those representing
nonlocal supersymmetry breaking. It is also of interest
to extend the analysis of \cite{peskin} to allow
one to match the higher dimensional theory directly to the four-dimensional
theory. Futhermore, a detailed understanding of the possibilities
of models derived from string theory, and whether they can
permit our assumptions would be important.
Further analysis of chiral theories derived from $D$-branes
in the presence of gravity is essential.

Finally, there is much to be better understood in the phenomenology.
One would want to include higher order loop corrections,
and also to perform the full renormalization group running. There
are interesting questions with respect to the viability of a large
tan beta fixed point in our scenario and a better understanding of the
implications of naturalness.

We find it encouraging that ultimately our scenario is testable.
There is little model dependence to our results; general features
of the spectrum reflect our underlying assumption
that the source of supersymmetry breaking resides in the
extra dimensions, sequestered from the visible world.

\bigskip
{\bf \noindent Acknowledgements:} We would like to thank
Jon Bagger, Andy Cohen, Sekhar Chivukula, 
Martin Gremm, Howie Haber, Petr Horava,
Burt Ovrut,  Erich Poppitz, Yael Shadmi,  Dan Waldram,
and Darien Wood for useful conversations. The research of Lisa Randall
was supported in part by DOE under cooperative
agreement DE-FC02-94ER40818 and
under grant number DE-FG02-91ER4071. The research of Raman Sundrum was 
supported by
the US Department of Energy under grant no. DE-FG02-94ER40818.

\section*{Appendix A: A Minimal Sequestered Sector}

In this Section, we give an example of a sequestered
sector. This illustrates how one treats the theory
from the low-energy four-dimensional vantage point. It 
will furthermore serve to illustrate the existence
of a nonzero VEV of $F_\phi$.

In this Appendix, we assume 
 that supersymmetry breaking is caused by
some strong dynamics in a hidden sector at an ``intermediate'' scale 
$\Lambda_H$, much larger than the weak scale. The details of the hidden
dynamics are irrelevant to  the weak-scale supersymmetry breaking effects
inherited by the visible sector. Here
we make the simplest assumption that  the only massless
state produced by the hidden sector is the Goldstino, and that all other
hidden sector states have typical masses of order $\Lambda_H$. (It would
however not significantly alter our analysis to include some extra massless
hidden sector states.) It therefore makes sense to imagine integrating
out all the massive (non-perturbative) $\Lambda_H$-scale physics. 
So we  replace the unknown (and presumably strongly
coupled and complicated) hidden sector by a simple model with a single
chiral superfield whose auxiliary component expecation value breaks
supersymmetry spontaneously at $\Lambda_H$.

We begin by switching off supergravity and considering a flat-superspace
hidden sector given by,
\begin{equation}
{\cal L} = \int d^4 \theta \Lambda_H^2 {\rm ln}(1 +
\frac{|\Sigma|^2}{\Lambda_H^2}) + \int d^2 \theta \Lambda_H^2 \Sigma
 + {\rm h.c.}
\end{equation}
We straightforwardly derive the scalar potential, 
\begin{equation}
{\cal V} = \Lambda_H^4 (1 + \frac{|\sigma|^2}{\Lambda_H^2})^2.
\end{equation}
We see immediately that,
\begin{eqnarray}
\langle \sigma \rangle &=& 0 \nonumber \\
m_{\sigma}^2 &=& 2 \Lambda_H^2 \nonumber \\
\langle F_{\Sigma} \rangle &=&  \Lambda_H^2 \nonumber \\
m_{\psi_{\Sigma}} &=& 0.
\end{eqnarray}
That is, supersymmetry is broken at the required scale and the only
massless state is the Goldstino. The scalar $\sigma$ is the analog of the
massive $\sigma$ mode of the linear $\sigma$ model. The
non-renormalizability of the model is unimportant since it is to be used
as a low-energy effective theory. All high-energy degrees of freedom
 have already been integrated out.  

Let us now couple just this sector to supergravity by taking 
eq. (\ref{lag}) to be given by,
\begin{eqnarray}
f &=& - 3 M_{Pl}^2 + \Lambda_H^2 {\rm ln} (1 + \frac{|\Sigma|^2}{\Lambda_H^2})
\nonumber\\
W &=& \Lambda_H^2(\Sigma + c), 
\end{eqnarray}
where the constant $c$ has been included to cancel the cosmological
constant. Our model now resembles the Polonyi model, but with a
non-canonical Kahler potential which gives the scalars $\sigma$ a
$\Lambda_H$-scale mass. Because the scalars are massive, the weak
gravitational couplings will only negligibly alter the VEVs found above for
$\Sigma$. We can use these VEVs in eq. (\ref{lag}) to solve for 
$\langle F_{\Phi} \rangle$. 
The $F_{\Phi}$ equation of motion simply yields, 
\begin{equation}
\langle F_{\Phi} \rangle  = \frac{\Lambda_H^2 c}{M_{Pl}^2}. 
\end{equation}
The vacuum energy following from the auxilary part of eq. (\ref{lag}) is
then given by, 
\begin{equation}
\langle {\cal V} \rangle  = \Lambda_H^4 (1 - 3 \frac{c^2}{M_{Pl}^2}). 
\end{equation}
Now the Weyl rescaling discussed in Section 4 would normally necessitate
multiplying the naive vacuum energy from eq. (\ref{lag}) by exp$(2
\langle K \rangle / 3 M_{Pl}^2$, but at $\sigma = 0$ one can see by
Eq. (\ref{Kdef}) that this factor is unity. The vacuum energy we have
computed is nothing but the  cosmological constant and must be tuned to
vanish. Thus we must have, 
\begin{eqnarray}
F_{\phi} &=& \sqrt{3} \frac{\Lambda^2_H}{M_{Pl}} \nonumber \\
c &=& \sqrt{3} M_{Pl}. 
\end{eqnarray}

The super-Higgs mechanism will cause the Goldstino to be eaten by the
gravitino, which will therefore aquires  a mass
Eq. ({\ref{gravitino}), 
\begin{equation}
m_{3/2} = \sqrt{3} \frac{\Lambda_H^2}{M_{Pl}}. 
\end{equation}

\section*{Appendix B: Regulating SQED}

Here we present a regularization for the simplest gauge theory, SQED, 
coupled to the supersymmetry-breaking background gravitational
superfield, $\Phi$. It will concretely illustrate
 how anomalous $\Phi$-dependence appears
after the rescaling of Eq. (\ref{rescaling}). A similar discussion appears in
Ref. \cite{analytic}. More general supersymmetric gauge theories are more
difficult to explicitly regulate and are treated in Ref. \cite{sweyl}.

The naive lagrangian for SQED coupled to the $\Phi$ background is
straightforwardly given by,
\begin{equation}
\label{sqed}
{\cal L}_{SQED} = \int d^4 \theta [Q^\dagger_+ e^V Q_+ |\phi|^2 + 
Q^\dagger_- e^V Q_- |\Phi|^2] + (\int d^2 \theta \frac{1}{g_0^2} {\cal
  W}^2 + {\rm h.c.}) + {\rm gauge\  fixing}.
\end{equation}
The fact that the gauge multiplet terms are $\Phi$-independent is a
reflection of the classical Weyl-invariance of their action. This
property is shared by the standard gauge-fixing terms, which we do not
write explicitly here. Of course the Feynman diagrams that follow from
this naive lagrangian are ultraviolet-divergent. We can
gauge-invariantly regulate closed loops of charged fields using the
time-honored Pauli-Villars procedure of adding massive charged regulator
fields with the ``wrong'' statistics, 
\begin{equation}
\label{pv}
 {\cal L}_{PV} = \int d^4 \theta [Q_+^{reg ~ \dagger} e^V Q^{reg}_+ 
|\Phi|^2 + Q^{reg ~ \dagger}_- e^V Q_-^{reg} |\phi|^2] + (\int d^2
\theta \Phi^3 \Lambda_{UV} Q_+^{reg} Q_-^{reg} + {\rm h.c.}).
\end{equation}
For sufficiently many such fields with oscillating statistics and masses
of order $\Lambda_{UV}$ any divergence from charged loops can be
regulated. For simplicity, we will retain just one pair of charged regulator
fields in this discussion. 

There will still be divergences in loops containing internal 
gauge-multiplet lines. For these we need to  soften the
behavior of the gauge propagators above the cutoff scale. For example,
we can regulate the (Feynman gauge) photon propagator as follows, 
\begin{equation}
\frac{\eta_{\mu \nu}}{p^2} \rightarrow \frac{\eta_{\mu \nu}}{p^2 (1 -
  p^2/\Lambda_{UV}^2)}.
\end{equation}
By multiplying by sufficiently many such cutoff ``form-factors'' we can
regulate any loops containing internal photon lines. Again for
simplicity we will just consider one such form-factor. The regulated
propagator  can be thought of as coming from a regulated lagrangian containing
higher derivatives, 
\begin{equation}
{\cal L}_{photon}^{reg} = -\frac{1}{4} F_{\mu \nu} (1 +
\frac{\partial^2}{\Lambda_{UV}^2}) F^{\mu \nu} + {\rm gauge-fixing}.
\end{equation}
While this is a satisfactory gauge-invariant regulated lagrangian, its
supersymmetrized form also involves higher derivatives and therefore
poses a needless complication. To avoid this we will separate the
regulated photon propagator into the difference of a naive propagator
and a massive propagator, 
\begin{equation} 
\frac{\eta_{\mu \nu}}{p^2 (1 - p^2/\Lambda_{UV}^2)} = 
\frac{\eta_{\mu \nu}}{p^2} - \frac{\eta_{\mu \nu}}{p^2 -
  \Lambda_{UV}^2}.
\end{equation}
The supersymmetrized gauge mass term is simply given by
\cite{wessbagger},
\begin{equation}
\delta {\cal L}_{gauge ~mass} = \int d^4 \theta \frac{1}{g_0^2} \Lambda_{UV}^2
V^2. 
\end{equation}
For an abelian gauge theory this mass term effectively does not break
gauge invariance since we can imagine it arising from the Higgs
mechanism, where the Higgs charge is infinitesimally small and the Higgs
VEV is infinitely large, in such a way
 that their product yields a $\Lambda_{UV}$ mass. In this limit, the
 physical Higgs degrees of freedom decouple and the Goldstone fields are
 eaten away, leaving only the gauge mass term. This trick cannot be used
 for non-abelian gauge fields because charge is quantized and therefore
 cannot be taken infinitesimal. This mechanism for giving mass to the
 gauge multiplet also makes it clear how $\Phi$-dependence must appear
 when coupled to supergravity. Since the mass term is effectively a
 conventional Higgs kinetic term in unitary gauge it must appear multiplied
 by $|\Phi|^2$ in the supergravity formalism. That is, 
\begin{equation}
\label{massive}
{\cal L} = \int d^4 \theta \frac{1}{g_0^2} \Lambda_{UV}^2
V^2 |\phi^2| + (\int d^2 \theta \frac{1}{g_0^2} {\cal
  W}^2 + {\rm h.c.}) + {\rm gauge-fixing}.
\end{equation} 

Let us then summarize the algorithm for regulating a general Feynman
supergraph: (i) Write the integral following from
Eq. (\ref{sqed}) plus Eq. (\ref{pv}). The latter's contributions will
regulate all charged loop (sub-)divergences. (ii) 
Subtract from the  naive super-propagator for each internal gauge line,
the super-propagator arising from the massive gauge lagrangian,
Eq. (\ref{massive}).

This procedure clarifies the claims made in  Section 6. In the naive
Lagrangian Eq. (\ref{sqed}), we can easily rescale away
$\Phi$-dependence altogether. This same rescaling can also be
applied to the Pauli-Villars regulator fields, but now we see that
$\Phi$-dependence is retained in the cutoff mass-term, 
\begin{equation}
\label{rescpv}
 {\cal L}_{PV} = \int d^4 \theta [Q_+^{reg ~ \dagger} e^V Q^{reg}_+ 
 + Q^{reg ~ \dagger}_- e^V Q_-^{reg}]  + (\int d^2
\theta \Phi \Lambda_{UV} Q_+^{reg} Q_-^{reg} + {\rm h.c.}).
\end{equation}
Similarly, the gauge multiplet regulator mass-squared is also
multiplied by $|\Phi|^2$. That is, the rescaled regulated theory depends
on $\Lambda_{UV}$ and $\Phi, \Phi^\dagger$, precisely through the
  combinations $\Lambda_{UV}\Phi, \Lambda_{UV}\Phi^\dagger$. Now let
  us consider the R-symmetry {\it before the rescaling} under which
  $R[\Phi] = 2/3$ and $R[Q] = 0$. This is clearly a valid symmetry even in the
    presence of the regulator. After the rescaling of course $R[Q] =
    2/3$. We still have a valid symmetry as long as we remember the
    $\Phi$-dependence appearing with the cutoff masses. If we simply go
    to the flat space theory now by imposing  $\Phi = 1$, then we must
    reproduce the fact that the R-symmetry is anomalous. We conclude
    that in the renormalized theory, anomalous $\Phi$ dependence must be
    present so as to cancel the anomaly that would otherwise be
    there. This only involves the phase of the lowest component of
    $\Phi$, but because $\Phi$ couples as a background superfield in 
 regulated SQED, the couplings of its various components after
 renormalization will be related by supersymmetry. More explicitly at
 one-loop and after rescaling, we can
 see that the usual logarithmic  divergence of the gauge vacuum
 polarization must come in the form $\int d^2 \theta$ 
ln$(\Lambda_{UV} \Phi) {\cal W}^2$ + h.c.,  since the
 regularization is purely due to Eq. (\ref{rescpv}), while the usual
 logarithmic divergence in the wavefunction of the charged fields must
 come in the form $\int d^4 \theta$ ln$(\Lambda_{UV}^2 |\Phi|^2) |Q|^2$,
 since the regularization is purely due to Eq. (\ref{massive}).

\end{document}